\documentclass[11pt,epic,misc]{article}
\usepackage{amssymb}
\usepackage{epsfig}
\usepackage{xcolor}
\usepackage{colortbl}

\def\Red{} 
\def\Blue{} 
\def\Black{} 

\catcode`\@=11
\def\numberbysection{\@addtoreset{equation}{section}
         \def\theequation{\thesection.\arabic{equation}}}
\numberbysection

\def\s{\sigma}

\def\o{\omega}
\def\D{\Delta}
\def\V{{\cal V}}
\def\R{{\cal R}}
\def\Gop{G^{\rm op}}
\def\Gcl{G^{\rm cl}}
\def\Dop{\D^{\rm op}}
\def\Dcl{\D^{\rm cl}}
\def\bi{\mathbb{I}}
\def\Z{\mathbb{Z}}
\def\la{\langle}
\def\ra{\rangle}
\renewcommand\L{\mathcal{L}}
\def\C{\mathcal C}
\def\lar{\leftarrow}

\definecolor{myyellow}{rgb}{.95,.96,.3}

\def\be{\begin{equation}}
\def\ee{\end{equation}}
\def\bea{\begin{eqnarray}}
\def\eea{\end{eqnarray}}
\newcommand\egal{&\!\!=\!\!&}

\begin{document}

\title{Wind on the boundary for the Abelian sandpile model}

\author{Philippe Ruelle\\
Institut de Physique Th\'eorique\\
Universit\'e catholique de Louvain\\ 
1348 Louvain-La-Neuve, Belgium}

\maketitle

\begin{abstract}
We continue our investigation of the two-dimensional Abelian sandpile model in terms of a logarithmic conformal field theory with central charge $c=-2$, by introducing two new boundary conditions. These have two unusual features: they carry an intrinsic orientation, and, more strangely, they cannot be imposed uniformly on a whole boundary (like the edge of a cylinder). They lead to seven new boundary condition changing fields, some of them being in highest weight representations (weights $-1/8,\,0$ and $3/8$), some others belonging to indecomposable representations with rank 2 Jordan cells (lowest weights 0 and 1). Their fusion algebra appears to be in full agreement with the fusion rules conjectured by Gaberdiel and Kausch. 
\end{abstract}

\section{Introduction}
After early appearances \cite{rosa,sal} and since its first formal study in \cite{gur}, much attention has been directed to logarithmic conformal field theories (LCFT). This has led to significant progress, both for abstract models (see the reviews \cite{flo,gab}) and for lattice realizations \cite{maru,gulu,ru,dgnpr,jeng1,piru1,jeng2,jeng3,piru2,piru3,marr,flomu,iprh, jpr,prz,pr,resa}. The known examples support the view that the critical lattice models described by logarithmic conformal theories have some intrinsic non-local features, in contrast to usual equilibrium models defined in terms of local Boltzmann weights with no global constraints. These non-local features on the lattice have an echo in the continuum in the form of peculiar properties of LCFTs, like the non-algebraic, logarithmic nature of their correlation functions, their non-unitarity, and non-diagonalizable Hamiltonians. 

Until very recently, only a few isolated lattice models were known to be describable by a LCFT. These include the dense polymer model \cite{sal,ivash,pr}, dilute polymers \cite{gulu}, percolation \cite{sadu,lpsa,car,wat,gulu,flomu,pr2}, the Abelian sandpile model (ASM) \cite{maru,ru,jeng1,piru1,jeng2,jeng3,piru2,piru3,marr,jpr} and the close-packed dimer model \cite{iprh}. The situation has considerably improved since an infinite family of integrable lattice models has been found \cite{prz,resa,pr3}, which correspond to  logarithmic extensions of the usual minimal models.

The simplest, and most studied LCFT has central charge $c=-2$. A local conformal field theory with this value of the central charge has been defined in \cite{gk,garu}, and is known as the triplet theory due to the existence of three conserved currents. This theory therefore possesses an extended algebra with respect to which it is rational. Despite the fact that the Virasoro representation theory for logarithmic theories has been worked out in some detail \cite{gk2,roh,eberflo}, no consistent, complete field theory with $c=-2$ other than the triplet theory has been constructed so far. 

On the lattice side, two models are known to have $c=-2$: the dense polymer model and the Abelian sandpile model (certain aspects of the dimer model can also be described by $c=-2$, see \cite{ipr}). Although none of them is described by the triplet theory, the question remains open however as to whether they fall in the same universality class, and are described by the same conformal theory.

An important question is therefore to investigate the precise structure of the logarithmic theories underlying these two lattice models. They are presumably more generic than the triplet theory which, because of its extended symmetry, is special. They are also likely to be quasi-rational but not rational. 

It is our purpose in this article to further investigate the nature of the theory describing the sandpile model. We do this by defining new boundary conditions, and by exploring their various properties with respect to each other (fusion rules) and with respect to other known observables. 


\section{The sandpile model}

We start with a brief description of the model \cite{btw}; for further details and more results, the reader is refered to \cite{dd}.

Every site $i$ of a rectangular grid $\L$ is assigned a height variable $h_i$, taking the four values $1, 2, 3$ and 4. A configuration $\C$ is the set of values $\{h_i\}$ for all sites. Given a configuration $\C_t$ at time $t$, the discrete dynamics makes it evolve to $\C_{t+1}$, in the following way. First the height variable of a random site $i$ is incremented by 1, $h_i \to h_i + 1$, making a new configuration $\C'_t$. Then if the (new) height $h_i$ in $\C'_t$ is smaller or equal to 4, one simply sets $\C_{t+1}=\C'_t$. If not, all sites $j$ such that their height variables $h_j$ are greater than 4 repeatedly topple, which means that $h_j$ is decreased by 4, whereas the height of all the nearest neighbours of $j$ are increased by 1. This toppling process stops when all height variables are between 1 and 4; the so obtained configuration defines $\C_{t+1}$.

The addition of a single sand grain may trigger a potentially very large avalanche: the toppling of one site transfers sand to its neighbours, which in turn can topple and so on. When the dynamics is run for long periods, the sandpile builds up, and is eventually subjected to an avalanche spanning the entire system. This correlates the height variables over very large distances, and makes the system critical in the thermodynamic limit.

The stochastic dynamics described above is Markovian and defines a time evolution in the space of probability distributions on the set of configurations. The asymptotic state of the sandpile is controlled by the unique invariant distribution $P_{\L}^*$. It can be shown that $P_\L^*$ is uniform on the set $\R$ of so-called recurrent configurations, and vanishes on the others. To be recurrent, a configuration must satisfy certain global conditions, and this is where non-locality enters. For what follows, it will be enough to know that the recurrent configurations are in one-to-one correspondence with the spanning trees on $\L$. 

More important is what happens on the boundaries. In the above description of the model, all boundary sites are dissipative: under toppling, since they have three or two nearest neighbours, one or two grains of sand fall off the pile. Such dissipative boundary sites are called open, and define the open boundary condition. The presence of dissipative sites is essential for the dynamics to be well-defined, since otherwise the toppling process might never stop. It is
convenient to think of the sand grains falling off the pile as being transferred to a
sink site, connected to all dissipative sites. 

Boundary sites can also be closed, i.e. non-dissipative. In this case, a toppling site loses as many sand grains as it has neighbours, generally three (two for corner sites). As a consequence, the height of a closed site takes only three (or two) values, usually chosen to be 1, 2 and 3 (1 and 2).

All boundary sites can be independently chosen to be open or closed, insofar as there is somewhere at least one dissipative site. If this is so, the dynamics is well-defined, and so is the set $\R$, which will depend on the boundary conditions chosen. The transfer of sand in a  toppling can be encoded in the toppling matrix $\D$,
\be
\D_{ij} = \cases{z_i & if $i=j$ is either a bulk site or a closed boundary
site,\cr
4 & if $i=j$ is an open boundary site,\cr
-1 & if $i,j$ are nearest neighbours, \cr
0 & otherwise,}
\label{top}
\ee
where $z_i$ is the number of neighbours of $i$. The matrix $\Delta$ is simply the discrete Laplacian on $\L$, subjected to specific boundary conditions.

In the infinite volume limit, the invariant measure $P^*_\L$ becomes a conformal field theoretic measure, and homogeneous open or closed boundaries are conformally invariant boundary conditions. In what follows, we will work on the upper-half plane (UHP), with an infinite-dimensional matrix $\D$, denoted by $\Dop$ resp. $\Dcl$ if the boundary is fully open resp. closed. The difference $\Dop -\Dcl$ is diagonal, with entries equal to 1 on boundary sites, and 0 elsewhere. 

These two boundary conditions, open and closed, have been well studied and are well understood (see \cite{ru,piru1,jpr} and below).


\section{Spanning trees}

As recurrent configurations on $\L$ are in correspondence with spanning trees on $\L^\star$, the grid $\L$ plus the sink site, Kirchhoff's theorem (see for instance \cite{priez}) implies that their number is given by
\be
|\R| = \det\D.
\ee

Spanning trees are acyclic configurations of arrows: at each site $i$ of $\L$, there is an outgoing arrow, pointing to any one of its $\D_{ii}$ neighbours (if $i$ is dissipative, the arrow can point to the sink site). A configuration of arrows defines a spanning tree if it contains no loop. By construction, the paths formed by the arrows all lead to the sink site, which is the root of the tree. If walked in the opposite direction, the paths start from the sink site (the root), enter the lattice through the open boundary sites, which are the only ones connected to the sink, ramify and eventually invade the lattice.

Even though the explicit mapping is very complicated and non-local, the spanning trees provide a description of recurrent configurations which is completely equivalent to that in terms of heights. The non-local conditions satisfied by the height values of recurrent configurations are encased in the property of arrow configurations to contain no loop, which is indeed a global constraint. However the spanning trees are often more convenient for actual calculations. In particular, they allow us to think of boundary conditions directly in terms of the arrows forming the spanning trees.

The simplest possibility is to force boundary sites to have their arrows pointing in the same direction, either left or right. This defines two new boundary conditions, which we denote by $\to$ and $\lar$. The more arrows we prescribe, and the stronger the wind blows in the trees ! The results below show that a wind blowing on the edge has long range effects deep in the forest and alters the general pattern of arrows far from the boundary. In what follows we examine the properties of the boundary fields which effect a change from an open or closed boundary condition to one of these two arrow boundary conditions, and in the limit where the finite grid goes to the upper-half plane (UHP). In order to review the way this is done, we start with the known case where an open boundary condition is changed to closed, and vice-versa \cite{ru}. 

Consider the sandpile model on the discrete UHP, with open boundary condition all along the boundary. It has toppling matrix $\Dop$. We modify the model slightly by closing one boundary site $i$. The new model has a slightly different toppling matrix which, from (\ref{top}), is simply given by $\D^{\rm new} = \Dop + B$, where the matrix $B$ is infinite-dimensional, but has a single non-zero entry, namely $B_{ii}=-1$. The determinants of $\Dop$ and $\D^{\rm new}$ give the number of recurrent configurations in both models, and are clearly infinite. However their ratio is finite, and is actually given by a 1-by-1 determinant (because $B$ has rank 1), where $\Gop = (\Dop)^{-1}$,
\be
{\det \D^{\rm new} \over \det \Dop} = \det (\bi + B \Gop) = 1 - \Gop_{ii} = {2 \over \pi} \simeq 0.637.
\label{1cl}
\ee
The number of recurrent configurations has decreased by a numerical (unimportant) factor, but otherwise, the closing of site $i$ has no noticeable effect at large distances. In the scaling limit, the two models are identical.

Things change if we close a large portion $I$ of an open boundary, say $n$ consecutive sites. Then the new model has toppling matrix $\D^{\rm new} = \Dop + B$, with $B$ equal to minus the identity matrix on $I$ and zero elsewhere. The ratio of determinants is now an $n$-by-$n$ determinant. It decays exponentially with $n$, like $e^{-2n{\rm G}/\pi}$ where G is the Catalan constant, because the effective free energy of a closed site is smaller than that of an open site by an amount equal to $2{\rm G}/\pi$. Subtracting this non-universal contribution from the free energy, a universal power law remains \cite{ru}
\be
e^{{2{\rm G} \over\pi}n} {\det \D^{\rm new} \over \det \Dop} = e^{{2{\rm G} \over\pi}n} \det (\bi + B \Gop)_{i,j \in I} \simeq A \, n^{1/4}\,, \qquad n \gg 1.
\label{opcl}
\ee
For $n$ large, the closing of $I$ has long range effects. In the scaling limit, the ratio of determinants can be viewed as the correlator of two boundary changing fields $\la \phi^{\rm op,cl}(0) \, \phi^{\rm cl,op}(n) \ra$, from which one concludes that the boundary condition changing field $\phi^{\rm op,cl} = \phi^{\rm cl,op}$ has dimension $-{1 \over 8}$.

The same procedure may be used for arrows. If we want to force the spanning trees to contain an arrow pointing from $i$ to $i+1$, two neighbour boundary sites, we modify two entries of the toppling matrix by setting $\D_{ii}=4+\delta$ and $\D_{i,i+1}=-1-\delta$. In the limit $\delta \to \infty$, $\det \D$ contains a term linear in $\delta$, whose coefficient yields exactly the number of spanning trees having the prescribed arrow. Consequently, we may define $\D^{\rm new} = \Dop + B$, where the two non-zero entries of $B$ are $B_{ii}=\delta$ and $B_{i,i+1}=-\delta$. Similar to (\ref{1cl}), the ratio 
\be
\lim_{\delta \to \infty} \; {1 \over \delta} {\det \D^{\rm new} \over \det \Dop} = \lim_{\delta \to \infty} \; {1 \over \delta} \det (\bi + B \Gop) = {3 \over 2} - {4 \over \pi} \simeq 0.227
\ee
gives the fraction of spanning trees which have an arrow between the sites $i$ and $i+1$. Exactly the same formula holds if an arrow is prescribed on a closed boundary, the only difference being in the use of the inverse Laplacian subjected to a closed boundary condition on the real axis. One finds
\be
\lim_{\delta \to \infty} \; {1 \over \delta} {\det \D^{\rm new} \over \det \Dcl} = \lim_{\delta \to \infty} \; {1 \over \delta} \det (\bi + B \Gcl) = {1 \over \pi} \simeq 0.318.
\ee
 
The same remarks as above can be made: it is only when a large number $n$ of arrows are forced on the boundary that long range effects develop. Prescribing $n$ consecutive arrows amounts to changing the entries of the toppling matrix through a perturbing matrix $B$, of dimension $n+1$. The determinant ratio has a term proportional to $\delta^n$, whose coefficient gives the fraction of trees having the $n$ prescribed arrows. As in the case of a change from open to closed, the asymptotic value of these determinant ratios can be evaluated exactly.


\section{Effect of the wind}

Suppose that we prescribe $n$ consecutive right arrows in an otherwise fully open or fully closed boundary (when there is only one string of arrows, their direction, left or right, makes no difference). The boundary itself is the set of points $\{(x,1) \in \Z^2\;:\; x \in \Z\}$; the first arrow points from $(1,1)$ to $(2,1)$, and the last $n$-th arrow from $(n,1)$ to $(n+1,1)$. We label these $n+1$ sites by an integer $1 \leq i \leq n+1$.

The defect matrix $B$ corresponding to this situation is zero except on this set of sites, where it equals \small
\be
B(\delta) = \pmatrix{\delta & -\delta & 0 & 0 & \cdots & 0\cr
 & \delta & -\delta &  &  &  \cr
& & \delta & -\delta &  & \cr
& & & \cdots & \cr
& & & & \delta & -\delta \cr
0 & 0 & 0 & \cdots & 0 & 0}.
\ee

\normalsize
The limit of $\delta^{-n} \det [\bi + B(\delta) G]$ for $\delta \to \infty$ reduces to the principal $(n+1,n+1)$ minor of $B(1) G$, and is explicitly given by\be
\lim_{\delta \to \infty} \; {1 \over \delta^n} {\det \D^{{\rm n}\to} \over \det \D} = \lim_{\delta \to \infty} {1 \over \delta^n} \det [\bi + B(\delta) G] = \det[G_{i,j} - G_{i+1,j}]_{1 \leq i,j \leq n},
\ee
where $G$ is $\Gop$ or $\Gcl$ depending on the boundary condition surrounding the arrows.
They are both invariant under horizontal translations, so that the previous determinant has the Toeplitz form $\det(\s_{i-j})$. Using the integral representation of the Green matrix on the plane and the method of images, one finds that the coefficients $\sigma_m = \int_0^{2\pi} {{\rm d}k \over 2\pi} e^{imk} \sigma(k)$ are the Fourier coefficients of a function $\sigma$, which explicitly reads
\be
\sigma^{\rm op}(k) = (1 - e^{ik}) \: \Big\{2 - \cos{k} - \sqrt{(3-\cos{k})(1-\cos{k})}\Big\},
\ee
for the open boundary condition, and 
\be
\sigma^{\rm cl}(k) = {1 \over 2} \, (1 - e^{ik}) \: \Big\{\sqrt{3-\cos{k} \over 1-\cos{k}} - 1\Big\},
\ee
for the closed boundary condition.

The asymptotic evaluation of Toeplitz determinants has been initiated in the celebrated Szeg\"o's theorem, and has since then been generalized in many directions. In the present case, the functions $\sigma^{\rm op}$ and $\sigma^{\rm cl}$ both contain a so-called Fisher-Hartwig singularity \cite{fishar}, for which the asymptotic analysis requires a fairly recent result by Ehrhardt and Silbermann \cite{ehrsi}. 

Let $\sigma(k)$ be a function on the unit circle with the following form, for ${\rm Re\ } \alpha > -{1 \over 2}$,
\be
\sigma(k) = e^{i\beta(k-\pi)} \, (2 - 2\cos{k})^\alpha \, \tau(k),
\label{fh}
\ee
where $\tau(k)$ is a smooth univalent function, nowhere vanishing nor divergent. Then the asymptotic value of the Toeplitz determinant formed with the Fourier coefficients of $\sigma$ is given by \cite{ehrsi}
\be
\det(\s_{i-j})_{1 \leq i,j \leq n} \simeq E[\s] \, n^{\alpha^2-\beta^2} \, e^{n(\log{\tau})_0}, \qquad n \gg 1.
\ee
In this expression, $(\log{\tau})_0$ is the zeroth Fourier coefficient of $\log{\tau}$, and $E[\s]$ is a constant whose explicit value can be found in \cite{ehrsi}. This remarkable theorem exactly fits the above situation since $\sigma^{\rm op}$ and $\sigma^{\rm ocl}$ have the form (\ref{fh}) with $\beta={1 \over 2}$, on account of
\be
1-e^{ik} = e^{i(k-\pi)/2} \, \sqrt{2-2\cos{k}}.
\ee
The other result, discussed in the previous section, about the closing of $n$ sites within an open boundary, follows from the same theorem, with $\beta=0$, a simpler case proved earlier by Widom \cite{widom}.

In case the boundary is open, from the function $\sigma^{\rm op}$, one has $\alpha=\beta={1 \over 2}$, and 
\be
\tau^{\rm op}(k) = 2 - \cos{k} - \sqrt{(3-\cos{k})(1-\cos{k})},
\ee
whose average value is equal to $(\log{\tau^{\rm op}})_0 = -{4{\rm G} \over \pi}$. We therefore obtain that, among all the recurrent configurations of the sandpile model on the open upper-half plane, the fraction of those which correspond to spanning trees having the $n$ prescribed arrows on the boundary, is equal, in the large $n$ limit, to
\be
{|\R^{{\rm op,}\to,{\rm op}}(n)| \over |\R^{\rm op}|} = E[\s^{\rm op}]\, e^{-{4{\rm G}\over \pi}n}.
\ee
The exponential decay has the same origin as the one discussed in the previous section, namely the fact that a site with a prescribed arrow has a smaller free energy than when the direction of the arrow is not constrained. Dividing out by this non-universal term, the ratio tends to a constant in the large $n$ regime,
\be
e^{{4{\rm G}\over \pi}n} \; {|\R^{{\rm op,}\to,{\rm op}}(n)| \over |\R^{\rm op}|} = E[\s^{\rm op}]\,, \qquad n \gg 1.
\label{open}
\ee

In the other case, when the boundary is closed, the function $\s^{\rm cl}$ has the form (\ref{fh}) with $\beta={1 \over 2},\, \alpha=0$, and 
\be
\tau^{\rm cl}(k) = \sqrt{3-\cos{k} \over 2} - \sqrt{1-\cos{k} \over 2}.
\ee
The exponential decay, controlled by $(\log{\tau^{\rm cl}})_0 = -{2{\rm G} \over \pi}$, is weaker than in the open case, as expected. However the main difference is that a universal power law remains since the ratio behaves, for large $n$, like
\be
e^{{2{\rm G}\over \pi}n} \; {|\R^{{\rm cl,}\to,{\rm cl}}(n)| \over |\R^{\rm cl}|} = E[\s^{\rm cl}]\,n^{-1/4}, \qquad n \gg 1.
\label{closed}
\ee


Let us now interpret these two results in the context of a conformal theory. The two cases involve the change of boundary conditions at two points, separated by a distance $n$. In conformal theory, these changes are effected by the insertion of specific boundary fields. Therefore the exponents which control the algebraic decay of the ratios of determinants, 0 and $1/4$ respectively for the open and closed boundary condition, should equal the sum of the dimensions of the related boundary condition changing fields. We will make the assumption that the potential field dimensions are contained in the (infinite) Kac table of the $(p,p')=(1,2)$ minimal model, which has $c=-2$, and given by
\be
h_{r,s} = {(2r-s)^2-1 \over 8}.
\ee
The values allowed by this formula are, in increasing order, $h=-{1 \over 8},\, 0,\, {3 \over 8},\, 1,\, {15 \over 8},\, 3, \ldots$ (part of the Kac table is displayed in Section 6). As the relevant conformal field theory is logarithmic \cite{jpr}, these scaling dimensions can specify either primary fields or logarithmic partners of primary fields.

For the open boundary condition, the sum of the dimensions of the two boundary fields is equal to 0, so that the dimension of both fields, $\phi^{{\rm op},\to}$ and $\phi^{\to,{\rm op}}$, must be 0. It is consistent to set, in the large $n$ limit,
\be
e^{{4{\rm G}\over \pi}n} \; {|\R^{{\rm op,}\to,{\rm op}}(n)| \over |\R^{\rm op}|} = \la \phi^{{\rm op},\to}(0) \phi^{\to,{\rm op}}(n) \ra = E[\s^{\rm op}].
\ee
One should note that, since the arrow boundary condition is oriented, the two fields $\phi^{{\rm op},\to}=\phi^{\lar,{\rm op}}$ (outgoing arrow) and $\phi^{{\rm op},\lar}=\phi^{\to,{\rm op}}$ (incoming arrow) are different. 

In the closed boundary case, the same argument leads to fields $\phi^{{\rm cl},\to}$ and $\phi^{\to,{\rm cl}}$ with dimensions adding up to $1 \over 4$. From the values of $h$ allowed by the Kac table, the only possibility is that one field has dimension $-{1 \over 8}$ and the other has dimension $3 \over 8$. This may look at first sight problematic because it prevents the identification of the ratio (\ref{closed}) with a 2-point correlator $\la \phi^{{\rm cl},\to}(0) \phi^{\to,{\rm cl}}(n) \ra$, as was done in the open case, since it vanishes identically. However one remembers that for the dynamics of the model to be well-defined, dissipation is essential. In the present case, because the whole boundary is non-dissipative (and there is no dissipation in the bulk either), one has to introduce dissipation at infinity by hand \cite{piru1,jpr}. It has been shown that the field corresponding to the insertion of an isolated dissipation, called $\omega_N$, is a dimension 0 field, logarithmic partner of the identity. Therefore the correct conformal interpretation of the result (\ref{closed}) is that it corresponds to a 3-point function,
\be
e^{{2{\rm G}\over \pi}n} \; {|\R^{{\rm cl,}\to,{\rm cl}}(n)| \over |\R^{\rm cl}|} = \la \phi^{{\rm cl},\to}(0) \, \phi^{\to,{\rm cl}}(n) \, \omega_N(\infty) \ra = E[\s^{\rm cl}]\,n^{-1/4}.
\ee

At this stage, by analyzing the change from open or closed to outgoing or incoming arrows, we have identified four new boundary condition changing fields: two fields $\phi^{{\rm op},\to}$ and $\phi^{\to,{\rm op}}$ with dimension 0, and two fields $\phi^{{\rm cl},\to}$ and $\phi^{\to,{\rm cl}}$, one with dimension $-{1 \over 8}$, the other with dimension $3 \over 8$, although we do not know yet which one is which. In addition to these four fields, there must be other fields which make the change from one arrow boundary condition to the other one, by reversing the direction of the arrows. We first determine their conformal weights. 


\section{Conformal weights}

In this section, we consider all changes of boundary conditions among open, closed, left arrows and right arrows, and determine the conformal weights of the corresponding fields. Under the assumption that these fields are primary, the conformal weights can be obtained rather easily by analyzing appropriate 3-point functions. It will be enough to consider an open boundary, in which we change the boundary condition on two consecutive intervals, say on $[0,L]$ and $[L,L+n]$. 

The effect of inserting two segments of different boundary conditions can be computed as before. The change of boundary condition on the two intervals is implemented by an appropriate defect matrix $B$, from which one computes the number of recurrent configurations (or spanning trees) for such a boundary as $\det[\Dop + B]$. Dividing that number by $\det \Dop$ yields the corresponding fraction of recurrent configurations as $\det[\bi + \Gop B]$, of dimension $L+n+1$. Since three changes of boundary condition are involved, the way this determinant behaves in the scaling limit should be given by 3-point functions, up to the usual exponential factors related to the boundary free energies. As we are assuming that the fields are (quasi-)primary, the general form of a 3-point correlator is
 \be
\la \phi_1(0) \, \phi_2(L) \, \phi_3(L+n) \ra = {\rm const.} \, {(L+n)^{h_2-h_1-h_3} \over L^{h_1+h_2-h_3} \: n^{h_2+h_3-h_1}}.
\label{3pt}
\ee

Because the determinants to be computed are no longer Toeplitz, the sort of theorem we used in the previous section for the insertion of a single segment cannot be applied, and we therefore resorted to numerical computations. The values of the determinants depend on $L$ and $n$. In all cases discussed below, the numerical calculations have been carried out for four fixed values of $L$, namely $L=20,\,30,\,50$ and 70, and for each value of $L$, $n$ was varied between $1$ and 150 (although the smallest values of $n$ were usually discarded in the analysis). By fitting the numerical data to the general form (\ref{3pt}), the values of the conformal weights can be extracted.

We first consider a string of $L$ arrows next to a segment of length $n$ of closed sites. This yields two different cases, according to whether the arrows point to the left or to the right. In both cases, the exponential factor by which we multiply the raw data is equal to $e^{(4L+2n){\rm G}/\pi}$, see (\ref{opcl}) and (\ref{open}).

For the arrows pointing to the right, we find that the numerical data are well reproduced by the above 3-correlator for the values of the scaling dimensions equal to $h_1=0$, $h_2={3 \over 8}$ and $h_3=-{1 \over 8}$. As the relevant 3-point function is 

\be
\setlength{\unitlength}{0.8mm}
\begin{picture}(70,0)(3,-1)
\multiput(0,0)(2,0){8}{\line(1,0){1}}
\put(15,-1){\line(0,1){2}} \Blue
\put(15,0){\vector(1,0){3}}
\put(18,0){\vector(1,0){3}}
\put(21,0){\vector(1,0){3}}
\put(24,0){\vector(1,0){3}}
\put(27,0){\vector(1,0){3}}
\put(30,0){\line(1,0){1}}\Red
\put(31,0){\line(1,0){18}}\Black
\put(31,-1){\line(0,1){2}}
\put(49,-1){\line(0,1){2}}
\multiput(49,0)(2,0){8}{\line(1,0){1}}
\put(7.5,-3){\makebox(0,0)[c]{\small op}}
\put(40.5,-2){\makebox(0,0)[c]{\small cl}}
\put(56.5,-3){\makebox(0,0)[c]{\small op}}
\put(23.5,2){\makebox(0,0)[b]{\small $L$}}
\put(40.5,2){\makebox(0,0)[b]{\small $n$}}
\end{picture}
\la \phi^{{\rm op},\to}(0) \, \phi^{\to,{\rm cl}}(L) \, \phi^{{\rm cl,op}}(L+n) \ra = {\rm const.} \, \Big({L+n \over L \sqrt{n}}\Big)^{1/2},
\ee
we may conclude that the field $\phi^{{\rm cl},\lar}$ changing a closed boundary condition into ingoing arrows has dimension $3 \over 8$. From this it follows that the field $\phi^{{\rm op},\to}$ transforming the open boundary condition into outgoing arrows cannot be the identity. 

This first result, combined with those of the previous section, implies that the field $\phi^{{\rm cl},\to}$ changing the closed condition into outgoing arrows must have dimension $-{1 \over 8}$. This can be confirmed by looking at the previous situation with the arrows pointing to the left. In that case the numerical data are consistent with the general 3-correlator for $h_1=0$, and $h_2=h_3=-{1 \over 8}$. We then have
\be
\setlength{\unitlength}{0.8mm}
\begin{picture}(70,0)(11,-1)
\multiput(0,0)(2,0){8}{\line(1,0){1}}
\put(15,-1){\line(0,1){2}}\Blue
\put(15,0){\line(1,0){2}}
\put(19,0){\vector(-1,0){3}}
\put(22,0){\vector(-1,0){3}}
\put(25,0){\vector(-1,0){3}}
\put(28,0){\vector(-1,0){3}}
\put(31,0){\vector(-1,0){3}}\Red
\put(31,0){\line(1,0){18}}\Black
\put(31,-1){\line(0,1){2}}
\put(49,-1){\line(0,1){2}}
\multiput(49,0)(2,0){8}{\line(1,0){1}}
\put(7.5,-3){\makebox(0,0)[c]{\small op}}
\put(40.5,-2){\makebox(0,0)[c]{\small cl}}
\put(56.5,-3){\makebox(0,0)[c]{\small op}}
\put(23.5,2){\makebox(0,0)[b]{\small $L$}}
\put(40.5,2){\makebox(0,0)[b]{\small $n$}}
\end{picture}
\la \phi^{{\rm op},\leftarrow}(0) \, \phi^{\leftarrow,{\rm cl}}(L) \, \phi^{{\rm cl,op}}(L+n) \ra = {\rm const.} \, n^{1/4},
\ee
which confirms that $\phi^{{\rm cl},\to}$ has indeed dimension $-{1 \over 8}$. Again the field $\phi^{{\rm op},\leftarrow}$ of dimension 0 cannot be the identity.

The last boundary condition changing fields to be identified are those which reverse the direction of arrows. There are two different cases: a change from right to left arrows, or from left to right arrows. We start with the first case.

Let us assume that the change from right arrows $\to$ to left arrows $\leftarrow$ takes place at the point $x$. On the lattice, this corresponds to a string of right arrows, the last one pointing to $x$, and then a string of left arrows, the first one also pointing to $x$. The site $x$ itself can either be open or closed. This makes a difference if one remembers that the arrows eventually lead to the root of the spanning tree through some dissipative sites. If the site $x$ is non-dissipative (closed), then the chains of arrows pointing to $x$ from either side, must go back into the interior of the lattice in order to find their way out to the root. But if $x$ is dissipative (open), then the arrows can go out to the root directly through $x$. The global pattern of the arrows making up the tree is then completely different in the two cases, so that the presence of dissipation, even concentrated at a single site, has strong effects. Therefore we distinguish two fields $\phi^{\to \stackrel{\rm op}{,}\leftarrow}$ and $\phi^{\to \stackrel{\rm cl}{,}\leftarrow}$ according to whether the point at which the transition takes place is dissipative or not. 

The scaling dimensions of these two fields can be determined by computing the determinants corresponding to the situation where an open boundary contains a string of right arrows immediately followed by a string of left arrows, of length $L$ and $n$ respectively, and where the transition site is either open or closed. The exponential factor related to the free energies is now equal to $e^{4(L+n){\rm G}/\pi}$. The values of the determinants multiplied by these exponentials, for $L,n$ in the same range as before, are consistent with the following correlators,
\be
\setlength{\unitlength}{0.8mm}
\begin{picture}(70,0)(13,-2)
\multiput(0,0)(2,0){8}{\line(1,0){1}}
\put(15,-1){\line(0,1){2}}\Blue
\put(15,0){\vector(1,0){3}}
\put(18,0){\vector(1,0){3}}
\put(21,0){\vector(1,0){3}}
\put(24,0){\vector(1,0){3}}
\put(27,0){\vector(1,0){3}}
\put(30,0){\line(1,0){2}}\Black
\put(32,0){\circle*{1.5}}\Blue
\put(34,0){\line(-1,0){2}}
\put(37,0){\vector(-1,0){3}}
\put(40,0){\vector(-1,0){3}}
\put(43,0){\vector(-1,0){3}}
\put(46,0){\vector(-1,0){3}}
\put(49,0){\vector(-1,0){3}}\Black
\put(49,-1){\line(0,1){2}}
\multiput(49,0)(2,0){8}{\line(1,0){1}}
\put(32,-3){\makebox(0,0)[c]{\small op}}
\put(7.5,-3){\makebox(0,0)[c]{\small op}}
\put(56.5,-3){\makebox(0,0)[c]{\small op}}
\put(23.5,2){\makebox(0,0)[b]{\small $L$}}
\put(40.5,2){\makebox(0,0)[b]{\small $n$}}
\end{picture}
\la \phi^{{\rm op},\to}(0) \, \phi^{\to\stackrel{\rm op}{,}\leftarrow}(L) \, \phi^{\leftarrow,\rm op}(L+n) \ra = {\rm const.},
\ee
\be
\setlength{\unitlength}{0.8mm}
\begin{picture}(70,0)(1.5,-1)
\multiput(0,0)(2,0){8}{\line(1,0){1}}\Blue
\put(15,-1){\line(0,1){2}}
\put(15,0){\vector(1,0){3}}
\put(18,0){\vector(1,0){3}}
\put(21,0){\vector(1,0){3}}
\put(24,0){\vector(1,0){3}}
\put(27,0){\vector(1,0){3}}
\put(30,0){\line(1,0){2}}
\put(34,0){\line(-1,0){2}}
\put(37,0){\vector(-1,0){3}}
\put(40,0){\vector(-1,0){3}}
\put(43,0){\vector(-1,0){3}}
\put(46,0){\vector(-1,0){3}}
\put(49,0){\vector(-1,0){3}}\Red
\put(32,0){\circle*{1.5}}\Black
\put(49,-1){\line(0,1){2}}
\multiput(49,0)(2,0){8}{\line(1,0){1}}
\put(32,-3){\makebox(0,0)[c]{\small cl}}
\put(7.5,-3){\makebox(0,0)[c]{\small op}}
\put(56.5,-3){\makebox(0,0)[c]{\small op}}
\put(23.5,2){\makebox(0,0)[b]{\small $L$}}
\put(40.5,2){\makebox(0,0)[b]{\small $n$}}
\end{picture}
\la \phi^{{\rm op},\to}(0) \, \phi^{\to\stackrel{\rm cl}{,}\leftarrow}(L) \, \phi^{\leftarrow,\rm op}(L+n) \ra = {\rm const.} \, {L+n \over L n}.
\ee
As $\phi^{{\rm op},\to}=\phi^{\leftarrow,{\rm op}}$ has weight 0, we conclude that the fields $\phi^{\to\stackrel{\rm op}{,}\leftarrow}$ and $\phi^{\to\stackrel{\rm cl}{,}\leftarrow}$ have dimension 0 and 1, respectively. 

For the second case, we consider the change of arrows from $\leftarrow$ to $\to$. The sites where the transition occurs are non-dissipative by construction (they have their arrow pointing along the boundary), so there is only one type of boundary condition changing field $\phi^{\leftarrow,\to}$. This time the numerical calculations are consistent with a 3-point function going to a constant for large $L$ and $n$, namely
\be
\setlength{\unitlength}{0.8mm}
\begin{picture}(65,0)(15,-1)
\multiput(0,0)(2,0){8}{\line(1,0){1}}
\put(15,-1){\line(0,1){2}}\Blue
\put(15,0){\line(1,0){1}}
\put(19,0){\vector(-1,0){3}}
\put(22,0){\vector(-1,0){3}}
\put(25,0){\vector(-1,0){3}}
\put(28,0){\vector(-1,0){3}}
\put(31,0){\vector(-1,0){3}}
\put(32,0){\circle*{1.5}}
\put(33,0){\vector(1,0){3}}
\put(36,0){\vector(1,0){3}}
\put(39,0){\vector(1,0){3}}
\put(42,0){\vector(1,0){3}}
\put(45,0){\vector(1,0){3}}\Black
\put(49,-1){\line(0,1){2}}
\multiput(48,0)(2,0){8}{\line(1,0){1}}
\put(7.5,-3){\makebox(0,0)[c]{\small op}}
\put(56.5,-3){\makebox(0,0)[c]{\small op}}
\put(23.5,2){\makebox(0,0)[b]{\small $L$}}
\put(40.5,2){\makebox(0,0)[b]{\small $n$}}
\end{picture}
\la \phi^{{\rm op},\leftarrow}(0) \, \phi^{\leftarrow,\to}(L) \, \phi^{\to,\rm op}(L+n) \ra = {\rm const.}
\ee
It implies that the boundary condition changing field $\phi^{\leftarrow,\to}$ has dimension 0.

Therefore, associated with the four boundary conditions open, closed, $\to$ and $\lar$, we have eight boundary condition changing fields: four of dimension 0, two of dimension $-{1 \over 8}$, one of dimension $3 \over 8$ and another one of dimension 1. 

We know the conformal weights of the fields but not their precise nature. We have assumed that they are primary, and therefore generate a highest weight representation of the Virasoro algebra. But this is not a complete statement in a logarithmic conformal theory, as we need to answer the following questions: do these highest weight representations stand on their own, and if so, with which type of reducibility properties, or else are they part of bigger indecomposable representations, and if so, which ones ?


\section{Virasoro representations and fusion}

In the section, we make a short review of the results obtained originally by Gaberdiel and Kausch \cite{gk2} and independently by Rohsiepe \cite{roh} regarding irreducible and certain indecomposable Virasoro representations for $c=-2$, and their fusion. These authors have considered a much larger class of models, namely the models $(1,t)$ with central charges $c = 13 - 6(t+t^{-1})$ for $t \geq 1$ integer, but in what follows, we focus exclusively on $c=-2$, (mostly) following the notations of \cite{gk2}.

The infinite Kac table is filled with conformal weights $h_{r,s}$ given by $h_{r,s}={(2r-s)^2-1 \over 8}$ for $r,s \geq 1$. The lower left portion of the table is displayed below, with $r$ running horizontally rightwards, and $s$ running vertically upwards. Every scaling dimension in this table appears an infinite number of times. One may check that the first two rows $s=1,2$, in yellow, contain every number exactly once, as does the first column with the exception of the weight 0, which appears twice.

With every $h$ in this list is associated a highest weight representation ${\cal M}_h$, also called a Verma module, built on a highest weight state $|h\ra$ satisfying $L_0|h\ra = h \, |h\ra$, and annihilated by the positive modes of the Virasoro algebra, $L_n|h\ra=0$ for all $n>0$. It generates the representation by the free action of the negative modes,
\be
{\cal M}_h = {\rm span}\{L_{-n_1}L_{-n_2} \cdots |h\ra \;:\; n_1,n_2,\ldots \geq 0\}.
\ee

\renewcommand{\arraystretch}{1.4}
\begin{table}[t]
\begin{center}
\tabcolsep12pt
\begin{tabular}{| c | c | c | c | c | c | c | c }
$\vdots$ & $\vdots$ & $\vdots$ & $\vdots$ & $\vdots$ & $\vdots$ & $\vdots$ & \\
\hline
$15 \over 8$ & ${3 \over 8}$ & $-{1 \over 8}$ & $3 \over 8$ & $15 \over 8$ & $35 \over 8$ & ${63 \over 8}$ & $\cdots$\\
\hline
$1$ & $0$ & $0$ & $1$ & $3$ & $6$ & $10$ & $\cdots$ \\
\hline
${3 \over 8}$ & $-{1 \over 8}$ & $3 \over 8$ & $15 \over 8$ & $35 \over 8$ & ${63 \over 8}$ & ${99 \over 8}$ & $\cdots$ \\
\hline
$0$ & $0$ & $1$ & $3$ & $6$ & $10$ & 15 & $\cdots$ \\
\hline
\rowcolor{myyellow} $-{1 \over 8}$ & $3 \over 8$ & $15 \over 8$ & $35 \over 8$ & ${63 \over 8}$ & ${99 \over 8}$ & $143 \over 8$ & $\cdots$ \\
\hline
\rowcolor{myyellow} $0$ & $1$ & $3$ & $6$ & $10$ & 15 & 21 & $\cdots$ \\
\hline
\end{tabular}
\end{center}
\end{table}

The algebraic structure of the Verma modules ${\cal M}_h$ has been completely determined in \cite{feifu}. Each Verma module ${\cal M}_h$ contains, in addition to $|h\ra$ itself,  an infinite number of highest weight vectors, called singular vectors. They generate submodules, which form a linear chain of nested subspaces. If $h=h_{r,s}$, the module ${\cal M}_{r,s} \equiv {\cal M}_{h_{r,s}}$ contains a singular vector at level $rs$ (value of $L_0-h_{r,s}$), which generates a submodule ${\cal M}_{r+s,s}$. Correspondingly one defines the quasi-rational representation ${\cal V}_{r,s} = {\cal M}_{r,s}/{\cal M}_{r+s,s}$ by setting to zero this singular vector at level $rs$.

When $r,s$ run over the Kac table, one obtains an infinite number of highest weight representations ${\cal V}_{r,s}$. They are not all distinct: ${\cal V}_{r,s} = {\cal V}_{r',s'}$ if and only if $h_{r,s}=h_{r',s'}$ and $rs=r's'$, implying that the Kac table contains each ${\cal V}_{r,s}$ a finite number of times (which depends on $r,s$). For instance, the representations $\V_{r,1}$ appear only once, but the representations $\V_{r,2}$ appear twice in the table since ${\cal V}_{r,2} = {\cal V}_{1,2r}$ for every $r \geq 1$.

The representations ${\cal V}_{r,s}$ for all pairs $(r,s)$ with the same value of $h=h_{r,s}$ have different reducibility properties. Indeed a value of $h_{r,s}$ is attained for an infinite number of pairs $(r,s)$, but with increasing values of $rs$. Therefore, the representations ${\cal V}_{r,s}$ become more and more reducible as one goes deeper in the Kac table: if $h_{r,s}=h_{r',s'}$ and $r's' > rs$, then ${\cal V}_{r',s'}$ is the quotient of ${\cal M}_{r',s'}={\cal M}_{r,s}$ by a singular vector lying at a higher level than for ${\cal V}_{r,s}$, and this implies the inclusion ${\cal V}_{r,s} \subset {\cal V}_{r',s'}$.

The only irreducible representations ${\cal V}_{r,s}$ are those for $s=1,2$ (and for $r=1$, $s$ even by the above equivalence). The corresponding Verma modules have nested submodules given by
\be
{\cal M}_{r,s} \longrightarrow {\cal M}_{r+s,s} \longrightarrow {\cal M}_{r+2s,s} \longrightarrow {\cal M}_{r+3s,s} \longrightarrow \cdots \qquad\quad (r \geq 1, \, s=1,2),
\ee
implying that the singular vectors appear respectively at level $rs, (2r+s)s,(3r+3s)s,\ldots$ The characters of these irreducible representations are 
\be
\chi^{\rm irr}_{r,s}(q) = {q^{{1 \over 12}+h_{r,s}} \over \prod_{n=1}^\infty (1-q^n)} \times (1 - q^{rs}), \qquad r \geq 1, s=1,2.
\ee

The representations ${\cal V}_{r,s}$ with $r,s$ in the Kac table exhaust the set of highest weight representations which are degenerate, that is, which contain a null vector at a finite level. With each of them is associated a primary field satisfying a degeneracy condition at level $rs$. This in turn implies that the correlators containing this primary field satisfy a partial differential equation of order $rs$. 

The previous discussion shows no essential difference with the usual treatment of minimal models. New features arise when one considers fusions. Indeed a general result of \cite{gk2,roh} is that the fusion of the highest weight representations $\V_{r,s}$ closes on a bigger set of representations, which not only contains the $\V_{r,s}$ themselves but also new, indecomposable representations $\R_{r,1}$. We briefly recall the basic structure of the representations $\R_{r,1}$ following \cite{gk2}. 

The indecomposable representations $\R_{r,1}$ can be seen as made up of two reducible highest weight  representations $\V_{1,2r-1}$ and $\V_{1,2r+1}$ ($r \geq 1$). The two representations are graphically pictured by two vertical lines, tied together by the action of the Virasoro algebra (arrows). The dots represent certain specific states, and crosses mark singular null vectors.

\setlength{\unitlength}{1mm}
\begin{picture}(30,70)(-5,-3)
\put(0,0){\circle*{1.5}}
\put(0,15){\circle*{1.5}}
\put(0,30){\makebox(0,0)[c]{\footnotesize $\times$}}
\put(0,45){\makebox(0,0)[c]{\footnotesize $\times$}}
\put(0,2){\vector(0,1){11}}
\put(0,17){\vector(0,1){11}}
\put(0,32){\vector(0,1){11}}
\put(0,47){\vector(0,1){5}}
\put(20,15){\circle*{1.5}}
\put(20,30){\circle*{1.5}}
\put(20,45){\makebox(0,0)[c]{\footnotesize $\times$}}
\put(20,17){\vector(0,1){11}}
\put(20,32){\vector(0,1){11}}
\put(20,47){\vector(0,1){5}}
\put(18,15){\vector(-1,0){16}}
\put(18,30){\vector(-1,0){16}}
\put(18,45){\vector(-1,0){16}}
\put(18.7,14){\vector(-4,-3){17}}
\put(18.7,29){\vector(-4,-3){17}}
\put(-4,0){\makebox(0,0)[c]{\small $\xi_r$}}
\put(-4,15){\makebox(0,0)[c]{\small $\phi_r$}}
\put(-4,30){\makebox(0,0)[c]{\small $\phi'_r$}}
\put(24,15){\makebox(0,0)[c]{\small $\psi_r$}}
\put(24,30){\makebox(0,0)[c]{\small $\rho_r$}}
\put(24,45){\makebox(0,0)[c]{\small $\rho'_r$}}
\put(0,57){\makebox(0,0)[c]{\small $\V_{1,2r-1}$}}
\put(20,57){\makebox(0,0)[c]{\small $\V_{1,2r+1}$}}
\end{picture}

\vskip -7.1truecm
\hangindent=4.5cm 
\hangafter=-14
For $r \geq 2$, the two highest weight states have different dimension. The left representation $\V_{1,2r-1}$ is a reducible highest weight subrepresentation within $\R_{r,1}$, with highest weight vector $\xi_r$. The first singular vector $\phi_r$ lies at level $r-1$, but is however not null in $\V_{1,2r-1}$. The second singular vector $\phi'_r$, at level $2r-1$, is null both in $\V_{1,2r-1}$ and in $\R_{r,1}$, so that $\phi_r$ generates an irreducible subrepresentation $\V_{r,1}$ of $\R_{r,1}$. The normalizations of $\phi_r,\phi'_r$ can be fixed by requiring
\be
\phi_r = (L_{-1}^{r-1} + \cdots) \xi_r\,, \qquad \phi'_r = (L_{-1}^r + \cdots) \phi_r \equiv 0\,.
\ee
The right representation $\V_{1,2r+1}$ has highest weight state $\psi_r$, with the same conformal weight as $\phi_r$. The descendants of $\psi_r$, represented on the vertical line above $\psi_r$, are obtained in the usual way by the action of the (strictly) negative Virasoro modes on $\psi_r$, to which descendants of $\xi_r$ are added in an appropriate way. Among them are $\rho_r$ and $\rho'_r$, respectively the first and second singular vectors of $\V_{1,2r+1}$, when this representation is taken alone. In $\R_{r,1}$, positive Virasoro modes map $\rho_r$ to descendants of $\phi_r$, and $\rho'_r$ to descendants of $\phi'_r$. Thus only $\rho'_r$ is singular in $\R_{r,1}$, and set to zero, $\rho'_r \equiv 0$.

Whereas $\V_{1,2r-1}$ is a subrepresentation, $\V_{1,2r+1}$ is not, as it generates states of $\V_{1,2r-1}$ when acted on by $L_0$ and the positive Virasoro modes (in addition to states of $\V_{1,2r+1}$ itself). This off-diagonal action, making the full representation $\R_{r,1}$ indecomposable, can be defined by
\be
L_0 \psi_r = h_{r,1} \psi_r + \phi_r\,, \qquad L_1^{r-1} \psi_r = \beta_r \, \xi_r\,, \qquad L_p \psi_r = 0\,, \quad {\rm for \ }p \geq 2.
\ee
They show that the full representation $\R_{r,1}$ can in fact be generated from the single (cyclic) state $\psi_r$, although itself not a highest weight vector. The parameter $\beta_r$ cannot be absorbed in the normalization of the various states, and so is an intrinsic parameter which labels inequivalent representations. 

The only states in $\R_{r,1}$ which are highest weight states (which correspond to primary fields) are $\xi_r$ and $\phi_r$. They generate subrepresentations $\V_{1,2r-1}$ and $\V_{r,1}$, respectively reducible and irreducible. The quotient representation $\R_{r,1}/\V_{1,2r-1}$ obtained by setting $\xi_r$ to zero, is equal to $\V_{1,2r+1}$. It may be noted that $\R_{r,1}$ also contains an indecomposable subrepresentation, generated by $\rho_r$, whose structure is equal to the quotient of $\R_{r+1,1}$ by its $\phi_{r+1}$ and $\rho_{r+1}$ states. Quotient representations will play an important role in the subsequent sections.

\setlength{\unitlength}{1mm}
\begin{picture}(30,70)(-5,-3)
\put(0,15){\circle*{1.5}}
\put(0,30){\makebox(0,0)[c]{\footnotesize $\times$}}
\put(0,45){\makebox(0,0)[c]{\footnotesize $\times$}}
\put(0,17){\vector(0,1){11}}
\put(0,32){\vector(0,1){11}}
\put(0,47){\vector(0,1){5}}
\put(20,15){\circle*{1.5}}
\put(20,30){\circle*{1.5}}
\put(20,45){\makebox(0,0)[c]{\footnotesize $\times$}}
\put(20,17){\vector(0,1){11}}
\put(20,32){\vector(0,1){11}}
\put(20,47){\vector(0,1){5}}
\put(18,15){\vector(-1,0){16}}
\put(18,30){\vector(-1,0){16}}
\put(18,45){\vector(-1,0){16}}
\put(18.7,29){\vector(-4,-3){17}}
\put(-4,15){\makebox(0,0)[c]{\small $\phi_1$}}
\put(-4,30){\makebox(0,0)[c]{\small $\phi'_1$}}
\put(24,15){\makebox(0,0)[c]{\small $\o_1$}}
\put(24,30){\makebox(0,0)[c]{\small $\rho_1$}}
\put(24,45){\makebox(0,0)[c]{\small $\rho'_1$}}
\put(0,57){\makebox(0,0)[c]{\small $\V_{1,1}$}}
\put(20,57){\makebox(0,0)[c]{\small $\V_{1,3}$}}
\end{picture}

\vskip -7.1truecm
\hangindent=4.5cm 
\hangafter=-11
For $r=1$, the highest weight states of $\V_{1,1}$ and $\V_{1,3}$ have both conformal dimension 0. The one on the left generates an irreducible subrepresentation: its first singular vector, at level 1, is set to zero, $\phi'_1 = L_{-1} \phi_1 = 0$, implying that $\phi_1$ is the identity field. The other representation on the right has also a lowest state $\o_1$ with conformal weight 0. The Virasoro modes act on it in a way similar to the case $r>1$, except that there is no state below $\phi_1$ to which $\o_1$ can be mapped. The defining relations are thus
\be
L_0 \o_1 = \phi_1\,, \qquad L_p \o_1 = 0\,, \quad {\rm for \ }p \geq 1.
\ee
Again the full representation $\R_{1,1}$ is generated from the single state $\o_1$. As for $r>1$, $\rho_1$ (here at level 1) is not singular in $\R_{1,1}$, but $\rho'_1$, at level 3, is singular in $\R_{r,1}$ and null. The only highest weight state of $\R_{1,1}$ is $\phi_1$, the corresponding quotient $\R_{1,1}/\V_{1,1}$ being equal to $\V_{1,3}$. The vector $\rho_1$ generates an indecomposable subrepresentation, which has the same structure as a quotient of $\R_{2,1}$.

According to \cite{gk2}, the set of representations $\V_{r,s}$ and $\R_{r,1}$ for $r \geq 1$ and $s=1,2$, is closed under fusion. Relying on explicit checks on the first levels, the authors of \cite{gk2} were led to a general conjecture regarding the fusion rules. As far as we know, this conjecture has not been proved, but has been verified in a number of examples, most notably in \cite{pr}, where reducible highest weight representations are also considered, and in \cite{resa}. The fusion rules read \cite{gk2} 
\bea
\V_{r_1,1} \star \V_{r_2,1} \egal \bigoplus_m \; \V_{m,1}, \qquad 
\V_{r_1,1} \star \V_{r_2,2} = \bigoplus_m \; \V_{m,2}, \\
\V_{r_1,2} \star \V_{r_2,2} \egal \bigoplus_m \; \R_{m,1}, \qquad 
\V_{r_1,1} \star \R_{r_2,1} = \bigoplus_m \; \R_{m,1},\\
\V_{r_1,2} \star \R_{r_2,1} \egal \bigoplus_m \; [\V_{m-1,2} \oplus 2\,\V_{m,2} \oplus \V_{m+1,2}], \\
\R_{r_1,1} \star \R_{r_2,1} \egal \bigoplus_m \; [\R_{m-1,1} \oplus 2\,\R_{m,1} \oplus \R_{m+1,2}],
\eea
where the summations are over $m = |r_1-r_2|+1,\, |r_1-r_2|+3, \ldots,\, r_1+r_2-1$, and $\V_{0,2}=\R_{0,1} \equiv 0$. Two remarks should be made regarding these fusions\footnote{I am grateful to Matthias Gaberdiel for stressing these.}. First, each indecomposable representation $\R_{r,1}$ arising in these fusion rules has a {\it definite and constant value of $\beta_r$}. The first few values of $\beta_r$ have been computed in \cite{gk2}, for instance $\beta_2=-1,\, \beta_3=-18$. Thus whenever the representation $\R_{2,1}$ appears above, it is meant to be the one with parameter $\beta_2=-1$. The second remark is that if a given representation arises in a fusion product, any of its quotient representations can also arise. 

Indecomposable representations of the above type have already appeared in the conformal description of the sandpile model, since it has been shown that the (boundary or bulk) dissipation field  belongs to the lowest level of a $\R_{1,1}$ representation \cite{piru1}, and that the bulk height fields are found on the first level of a (non-chiral) $\R_{2,1}$ representation \cite{jpr}.


\section{Field identifications}

In this section, we derive more constraints on the nature of the boundary fields found in Section 5. Our analysis and proposal will be based on two main assumptions: (i) the boundary condition changing fields are primary, and degenerate at an as low level as possible, and (ii) they belong to representations $\V_{r,s}$ or $\R_{r,1}$ reviewed in the previous section, or to a quotient thereof. Assumption (i) is a consequence of the usual interpretation of the lowest lying state of the Hilbert space ${\cal H}_{\alpha,\beta}$ as corresponding to the field changing the boundary condition from $\alpha$ to $\beta$. There is no deep reason for the second assumption, except that the $\V_{r,s}$ and $\R_{r,1}$ appear to form a natural supply of representations. Let us note that, as a consequence, primary fields with fractional dimension necessarily belong to irreducible highest weight representations $\V_{r,2}$. In the next section, a few checks will be presented that support these assumptions.

We will not dwell much on the field that swaps the open and closed boundary conditions. It has already been much discussed in \cite{ru,piru1,jpr}, where considerable evidence shows that this field, called $\mu$, is a primary field of scaling dimension $-{1 \over 8}$, and corresponds to the highest weight state of an irreducible representation $\V_{1,2}$, degenerate at level 2. 

Apart from the field $\mu$, the two easy cases are the other two fields with dimension $-{1 \over 8}$ and $3 \over 8$, which we will call respectively $\mu'$ and $\nu$. For them, our two assumptions above lead directly to the conclusion that they are the highest weight states of the irreducible representations $\V_{1,2}$ and $\V_{2,2}$, degenerate respectively at level 2 and 4. 

All other fields have integral dimensions. To determine their properties, we mainly use the constraints coming from their composition law, expressed by the fusion algebra: for any boundary condition $\gamma$, the fusion $\phi^{\alpha,\gamma} \star \phi^{\gamma,\beta}$ must close on fields which interpolate between the boundary conditions $\alpha$ and $\beta$, and should in particular contain the boundary condition changing field $\phi^{\alpha,\beta}$.

Then from the fusion
\be
\phi^{\rm op,cl} \star \phi^{\rm cl,\to} = \mu \star \mu' = \V_{1,2} \star \V_{1,2} = \R_{1,1},
\ee
we readily conclude that $\phi^{\rm op,\to}$ must belong to a representation $\R_{1,1}$. Being of dimension zero and different of the identity, it can only correspond to the $\o_1$ field in $\R_{1,1}$. However its primary partner $\phi_1$, namely the identity, decouples completely from all correlation functions since the identity in fact does not interpolate between two different boundary conditions. Stated differently, conformal transformations of a general correlator $\la \ldots \phi^{\alpha,{\rm op}}(z_1) \,\phi^{\rm op,\to}(z_2) \, \phi^{\to,\beta}(z_3) \ldots\ra$ will generate an inhomogeneous term 
$\la \ldots \phi^{\alpha,{\rm op}}(z_1) \, \phi^{\to,\beta}(z_3) \ldots\ra$ which vanishes identically because of the mismatch of boundary conditions on $[z_1,z_3]$. Thus the primary field $\phi_1$ is null, and $\phi^{\rm op,\to}$ becomes the highest weight state of $\V_{1,3} = \R_{1,1}/\V_{1,1}$ (as we shall see in the next section, the singular state at level 1 in $\V_{1,3}$ is not null; this makes sure that $\phi^{\rm op,\to}$ is not the identity). We call this field $\sigma$, leaving the notation $\omega$ for the logarithmic partner of the identity. 

Likewise the fusion 
\be
\phi^{\rm op,cl} \star \phi^{\rm cl,\lar} = \mu \star \nu = \V_{1,2} \star \V_{2,2} = \R_{2,1}
\ee
implies that $\phi^{\rm op,\lar}$ should be identified with the $\xi_2$ field in a $\R_{2,1}$ representation. It cannot descend to the quotient by $\phi_2=L_{-1}\xi_2$ since this would mean that $\xi_2$ itself is the identity. We set $\phi^{\rm op,\lar} \equiv \xi_2$.

Next case is $\phi^{\lar,\to}$ which can be seen as arising in the fusion
\be
\phi^{\lar,\rm cl} \star \phi^{\rm cl,\to} = \mu' \star \mu' = \V_{1,2} \star \V_{1,2} = \R_{1,1}.
\ee
The same arguments as for $\phi^{\rm op,\to}$ show that $\phi^{\lar,\to}$ should be identified with the $\o_1$ field of $\R_{1,1}$, projected down to the quotient $\V_{1,3}$. Let us note that this is consistent with the fact that $\phi^{\lar,\to}$ should also appear in $\phi^{\lar,\rm op} \star \phi^{\rm op,\to}$, equal to $\V_{1,3} \star \V_{1,3} = \V_{1,1} \oplus \V_{1,3} \oplus \V_{1,5}$ \cite{pr}. We set $\phi^{\lar,\to} \equiv \sigma'$. 

The field $\phi^{\to\stackrel{\rm cl}{,}\leftarrow}$, of dimension 1, must be in
\be
\phi^{\to,\rm cl} \star \phi^{\rm cl,\lar} = \nu \star \nu = \V_{2,2} \star \V_{2,2} = \R_{1,1} \oplus \R_{3,1}.
\ee
The only primary field with that dimension is the field $\xi_3$ in $\R_{3,1}$. So we set $\phi^{\to\stackrel{\rm cl}{,}\leftarrow} \equiv \xi_3$.

The last field to consider is $\phi^{\to\stackrel{\rm op}{,}\leftarrow}$, of dimension 0. The only fusion to contain it is
\be
\phi^{\to,\rm op} \star \phi^{\rm op,\lar} = \xi_2 \star \xi_2 = \R_{2,1} \star \R_{2,1} = 2\R_{1,1} \oplus 2\R_{2,1} \oplus 2\R_{3,1} \oplus \R_{4,1}.
\ee
In the representations appearing in this fusion, there are only two candidates for a primary field of weight 0: the field $\xi_2$ in a $\R_{2,1}$, or the field $\o_1$ in a $\R_{1,1}$ quotiented to a $\V_{1,3}$. However another consistency condition is that the fusion $\phi^{\rm op,\to} \star \phi^{\to\stackrel{\rm op}{,}\leftarrow}$ should close on  
the representation $\R_{2,1}$ containing $\phi^{\rm op,\lar}$. This rules out the second possibility if one uses $\V_{1,3} \star \V_{1,3} = \V_{1,1} \oplus \V_{1,3} \oplus \V_{1,5}$ \cite{pr}. One concludes that $\phi^{\to\stackrel{\rm op}{,}\leftarrow} \equiv \xi_2'$ must be the lowest lying state in a $\R_{2,1}$ representation. 

Table 1 summarizes the physical interpretation of these fields, along with their scaling dimension and the type of representation they belong to. We add two remarks.

\bigskip
\renewcommand{\arraystretch}{1.8}
\begin{table}[t]
\begin{center}
\tabcolsep6pt
\begin{tabular}{|c||c|c|c|c|}
\hline
$\phi^{\alpha,\beta}$ & open & closed & $\to$ & $\lar$ \\
\hline\hline
open & id. & \ $\mu=[-{1 \over 8}] \in \V_{1,2}$\ & $\sigma=[0] \in \V_{1,3}$ & $\xi_2=[0] \in \R_{2,1}$\\
\hline
closed & \ $\mu=[-{1 \over 8}] \in \V_{1,2}$\ & id. &  $\mu'=[-{1 \over 8}] \in \V_{1,2}$\ &\ $\nu=[{3 \over 8}] \in \V_{2,2}$\ \\
\hline
$\to$ &  $\xi_2=[0] \in \R_{2,1}$ & $\nu=[{3 \over 8}] \in \V_{2,2}$ & id. & 
$\renewcommand{\arraystretch}{1.2}
\begin{array}{c}
\xi_2'=[0] \in \R_{2,1} \ ({\rm center\  op})\\
\xi^{}_3=[1] \in \R_{3,1} \ ({\rm center\  cl})
\end{array}$ \\
\hline
$\lar$ & $\sigma=[0] \in \V_{1,3}$ & $\mu'=[-{1 \over 8}] \in \V_{1,2}$ & $\sigma'=[0] \in \V_{1,3}$ & id. \\
\hline
\end{tabular}
\end{center}
\caption{Summary of the fields $\phi^{\alpha,\beta}$ which make a change of boundary condition from $\alpha$ (row label) to $\beta$ (column). The numbers in square brackets denote the scaling dimensions.}
\end{table}

\smallskip \noindent
\underline{Remark 1}: The four dimension 0 fields $\sigma,\sigma',\xi_2$ and $\xi_2'$, although members of different representations, all satisfy the same degeneracy condition at level 3, since they are annihilated by $(L_{-1}^2-2L_{-2})L_{-1}$.

\medskip \noindent
\underline{Remark 2}: From the way they appear in fusions, the two indecomposable representations $\R_{2,1}$ and $\R_{3,1}$ found above should have parameters $\beta_2=-1$ and $\beta_3=-18$. The two fields $\xi^{}_2=\phi^{\rm op,\lar}$ and $\xi_2'=\phi^{\to\stackrel{\rm op}{,}\leftarrow}$ are primary with zero dimension and degenerate at level 3. Both have a singular descendant at level 1, which however cannot be null states since that would identify $\xi^{}_2$ and $\xi_2'$ with the identity. So the degeneracy level of $\xi^{}_2,\xi_2'$ is as low as possible. In contrast, this is not the case of $\xi_3 = \phi^{\to\stackrel{\rm cl}{,}\leftarrow}$. In a fully-fledged $\R_{3,1}$, the field $\xi_3$ is degenerate at level 5, but possesses the singular descendant $\phi_3$ at level 2. If the latter is set to zero, then $\xi_3$ would belong to a quotient representation $\R_{3,1}/\phi_3$ with two peculiarities. First, the parameter $\beta_3$ loses its meaning, namely it can be reabsorbed in the normalizations of the non-zero fields. And second, in the quotient, the state $\psi_3$ and all its descendants no longer have partners, since all the states of which they were the logarithmic partners have been set to zero. So the quotient is an indecomposable non-logarithmic representation. Following \cite{pr2}, this type of representation can be called reducible but indecomposable representations of rank 1. 

\medskip
One can now check that the physical interpretation we give to these fields is compatible with the way they should compose under fusion. For instance
\be
\nu = \phi^{\rm cl,\lar}=\V_{2,2} \;\in\; \phi^{\rm cl,\to} \star \phi^{\to\stackrel{\rm cl}{,}\leftarrow} = \mu' \star \xi_3 = \V_{1,2} \star \R_{3,1} = \V_{2,2} \oplus 2\V_{3,2} \oplus \V_{4,2},
\ee
and, using the results of \cite{pr},
\be
\xi_2 = \phi^{\rm op,\lar}=\R_{2,1} \;\in\; \phi^{\rm op,\to} \star \phi^{\to\stackrel{\rm op}{,}\leftarrow} = \sigma \star \xi'_2 = \V_{1,3} \star \R_{2,1} = \R_{1,1} \oplus \R_{2,1} \oplus \R_{3,1}.
\ee

In other cases, the composition law is satisfied provided one takes a quotient (see the second remark at the end of the previous section). Examples have been already encountered above, and another simple instance is provided by 
\be
\phi^{\rm op,cl} \star \phi^{\rm cl,op} = \mu \star \mu = \V_{1,2} \star \V_{1,2} = \R_{1,1}.
\ee
This fusion contains the identity in the double quotient of $\R_{1,1}$ by $\phi_1$ and $\rho_1$, since $\R_{1,1}/\{\phi_1,\rho_1\} = \V_{1,1}$. Despite the fact that $\R_{1,1}$ by itself contains the identity, the double quotient is necessary because the OPE $\mu(z) \mu(w)$ closes on an irreducible representation, as shown in \cite{piru1}.

Another example is
\be
\phi^{\rm cl,\to} \star \phi^{\to,\rm cl} = \mu \star \nu = \V_{1,2} \star \V_{2,2} = \R_{2,1},
\ee
also expected to contain the identity. However no quotient of $\R_{2,1}$ is equal to the irreducible $\V_{1,1}$. The only possibility is to quotient $\R_{2,1}$ by the singular state $\phi_2=L_{-1}\xi_2$, which forces the identification of $\xi_2$ with the identity. This quotient $\R_{2,1}/\phi_2$ is of the type discussed in the second remark above, indecomposable but non-logarithmic.


\section{Higher correlators}

In this section, we analyze higher correlators, and bring further support to the field identifications made in the previous section. We have investigated about a dozen 4-point amplitudes on the upper-half plane, but we merely present here four representative examples. They would in principle provide a check on the parameters $\beta_2,\beta_3$ for the representations $\R_{2,1}$ and $\R_{3,1}$, but for reasons explained below, no value could be extracted.

\subsection{\protect\boldmath Amplitudes with $\R_{2,1}$} 

There are a few calculable amplitudes which allow to study the fusion of $\V_{1,2} \star \V_{2,2} = \R_{2,1}$. One of them corresponds to the following situation. We consider the sandpile model on the UHP with an open boundary, in which we close the sites on a first segment, then force a string of right arrows on another consecutive segment, and finally close a third consecutive segment. It involves four changes of boundary conditions, 
say at $z_1 < z_2 < z_3 < z_4$, and corresponds to the following 4-point function,
\be
\setlength{\unitlength}{1mm}
\begin{picture}(80,7)(0,-1)
\multiput(0,0)(2,0){8}{\line(1,0){1}}
\put(15,-1){\line(0,1){2}}\Blue
\put(32,0){\vector(1,0){3}}
\put(35,0){\vector(1,0){3}}
\put(38,0){\vector(1,0){3}}
\put(41,0){\vector(1,0){3}}
\put(44,0){\vector(1,0){3}}\Black
\put(32,-1){\line(0,1){2}}
\put(47,-1){\line(0,1){2}}\Red
\put(47,0){\line(1,0){17}}
\put(15,0){\line(1,0){17}}\Black
\put(64,-1){\line(0,1){2}}
\put(23.5,-2){\makebox(0,0)[c]{\small cl}}
\put(55.5,-2){\makebox(0,0)[c]{\small cl}}
\multiput(64,0)(2,0){8}{\line(1,0){1}}
\put(71.5,-2.5){\makebox(0,0)[c]{\small op}}
\put(7.5,-2.5){\makebox(0,0)[c]{\small op}}
\put(15,3){\makebox(0,0)[c]{\small $z_1$}}
\put(32,3){\makebox(0,0)[c]{\small $z_2$}}
\put(47,3){\makebox(0,0)[c]{\small $z_3$}}
\put(64,3){\makebox(0,0)[c]{\small $z_4$}}
\end{picture}
\quad \sim \quad
\la \mu(z_1) \, \mu'(z_2) \, \nu(z_3) \, \mu(z_4) \ra.
\label{first}
\ee

From the degeneracy of $\mu$ at level 2, we obtain the general form of the above 4-point correlator, where $x={z_{12}z_{34} \over z_{13}z_{24}}$,
\be
\la \mu(z_1) \, \mu'(z_2) \, \nu(z_3) \, \mu(z_4) \ra = \Big({z_{12}z_{14}z_{24} \over z_{13}z_{23}z_{34}}\Big)^{1/4} \; \Big[ \alpha + \beta \log{x \over 1-x}\Big].
\ee
When $z_{12} \to 0$, the fusion of $\mu$ and $\mu'$ should close on the primary field $\s$ with no logarithmic singularity, implying $\beta=0$.

In order to compare with lattice results, we have taken the lengths of the two closed segments to be $L=z_{21}=z_{43}$, and that of the arrow interval to be $n=z_{32}$. In this situation, the variable $x$ is equal to $({L \over L+n})^2=t^2$, in terms of which the 4-point function reads
\be
\la \mu(0) \, \mu'(L) \, \nu(L+n) \, \mu(2L+n) \ra = \alpha \Big({2L+n \over n}\Big)^{1/4} = \alpha \Big({1+\sqrt{t} \over 1-\sqrt{t}}\Big)^{1/4}.
\label{fig1}
\ee

In actual calculations, we have fixed $L=20,\,30,\,50$ and 70, and for each value of $L$, $n$ has been varied between 1 and 150. The corresponding ratios of determinants have been numerically computed, and then divided by the appropriate exponential factors, equal here to $e^{-4(L+n)G/\pi}$. The previous conformal result suggests that the data, if plotted against the variable $t={L \over L+n}$, should collapse on a single curve. The data collapse is manifest in Figure 1, where the quantity
\be
e^{4(L+n)G/\pi} \; {\det \D^{\rm new} \over \det \Dop} 
\label{fig1latt}
\ee
is plotted as colour dots, in terms of the variable $t$. The solid black curve represents the function of $t$ in the r.h.s. of (\ref{fig1}), and for a value of the coefficient $\alpha$ fitted to the data ($\alpha \simeq 1.068$). This strongly supports the conclusion that indeed
\be
\la \mu(z_1) \, \mu'(z_2) \, \nu(z_3) \, \mu(z_4) \ra = \alpha \, \Big({z_{12}z_{14}z_{24} \over z_{13}z_{23}z_{34}}\Big)^{1/4}.
\ee

\begin{figure}[t]
\begin{center}
\epsfig{file=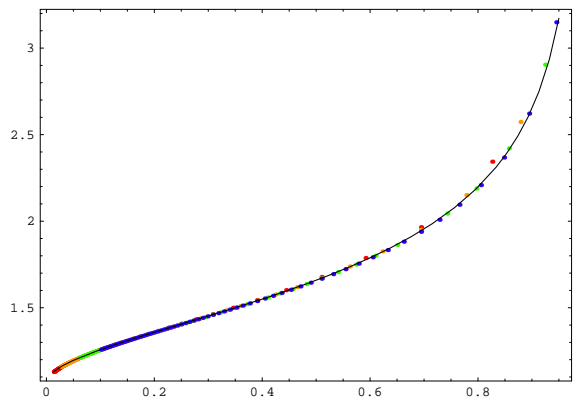,width = 7.5cm}
\end{center}
\caption{Comparison of numerical results against the conformal prediction for the situations depicted in (\ref{first}). The colour dots represent the lattice numerical results for (\ref{fig1latt}), for $n$ from 1 up to 150 and $L=20$ (red), 30 (orange), 50 (green) and 70 (blue). The abscissa is $t={L \over L+n}$.}
\end{figure}

When $z_3 \to z_4$, it can be reduced to a sum of 3-point functions upon using the OPE of $\mu$ and $\nu$. Conformal invariance fixes its form to be
\bea
\nu(z) \mu(0) \egal C_{\nu\mu}^{\R_{2,1}} \, z^{-1/4} \Big\{ \xi_2(0) + {z \over 2\beta_2} \log{z} \, \phi_2(0) + {z \over 2\beta_2} \, \psi_2(0) + {3z^2 \over 8\beta_2} \log{z} \, (L_{-1}\phi_2)(0) \nonumber\\
&& \hspace{1cm} -{3\beta_2+8 \over 16\beta_2} z^2 (L_{-1}\phi_2)(0) + {z^2 \over 4} \, (L_{-2}\xi_2)(0) + {3z^2 \over 8\beta_2} \, (L_{-1}\psi_2)(0) + \ldots \Big\}
\label{numu}
\eea
where $\beta_2$ is the parameter associated with the representation $\R_{2,1}$. 

Expanding the 4-point amplitude in powers of $z_{34}$, and using the previous OPE leads to the following 3-point functions
\bea
&& C_{\nu\mu}^{\R_{2,1}} \, \la \mu(z_1) \mu'(z_2) \xi_2(z_4) \ra = \alpha z_{12}^{1/4}\,, \qquad
C_{\nu\mu}^{\R_{2,1}} \, \la \mu(z_1) \mu'(z_2) \phi_2(z_4) \ra = 0, \\
\noalign{\medskip}
&& C_{\nu\mu}^{\R_{2,1}} \, \la \mu(z_1) \mu'(z_2) \psi_2(z_4) \ra = 
{\alpha \beta_2 \over 2} z_{12}^{1/4}\Big({1 \over z_{14}} + {1 \over z_{24}}\Big).
\eea
The first one has the correct form for primary fields of dimension $-{1 \over 8}$, $-{1 \over 8}$ and 0. The second and third ones are compatible with the first one and the relations $\phi_2=L_{-1}\xi_2$ and $L_1 \psi_2 = \beta_2 \xi_2$. Unfortunately the vanishing of the $\la \mu \mu' \phi_2\ra$ prevents us to check that $\beta_2$ has the expected value $-1$. We note that the expansion of the third relation around $z_1=z_2$ implies that $\sigma$ and its level 1 singular descendant $L_{-1}\sigma$ both have a non-zero 2-point function with $\psi_2$. This shows that $L_{-1}\sigma$ is not null, as announced in the previous section.

If one takes the limit $z_2 \to z_3$, the expansion in powers of $z_{23}$ yields similar results. However in this case, the primary field $\xi$ is expected to be the identity, so that its descendant $\phi=L_{-1}\xi$ clearly decouples since $\phi$ becomes null. 

Others situations allowing to study the OPE of $\V_{1,2}$ and $\V_{2,2}$ can be examined in the same way, with similar results (no logarithmic term, decoupling of the $\phi_2$ field). Among them, the following situation involves the boundary field related to having a height equal to 1 at an open boundary site. We consider again the UHP with an open boundary, in which we insert a segment of right arrows, close another segment of sites and ask that a given site in the remaining open boundary have a height variable equal to 1. 
In field theoretic terms, it corresponds to
\be
\setlength{\unitlength}{1mm}
\begin{picture}(80,7)(0,-1)
\multiput(0,0)(2,0){8}{\line(1,0){1}}
\put(15,-1){\line(0,1){2}}\Blue
\put(15,0){\vector(1,0){3}}
\put(18,0){\vector(1,0){3}}
\put(21,0){\vector(1,0){3}}
\put(24,0){\vector(1,0){3}}
\put(27,0){\vector(1,0){3}}
\put(30,0){\line(1,0){2}}\Black
\put(32,-1){\line(0,1){2}}\Red
\put(32,0){\line(1,0){15}}\Black
\put(47,-1){\line(0,1){2}}
\put(39.5,-2){\makebox(0,0)[c]{\small cl}}
\put(64,0){\circle*{1.5}}
\multiput(47,0)(2,0){17}{\line(1,0){1}}
\put(76,-2.5){\makebox(0,0)[c]{\small op}}
\put(7.5,-2.5){\makebox(0,0)[c]{\small op}}
\put(15,3){\makebox(0,0)[c]{\small $z_1$}}
\put(32,3){\makebox(0,0)[c]{\small $z_2$}}
\put(47,3){\makebox(0,0)[c]{\small $z_3$}}
\put(64,-3){\makebox(0,0)[c]{\small $z_4$}}
\put(64,3){\makebox(0,0)[c]{\small $h=1$}}
\end{picture}
\quad \sim \quad
\la \sigma(z_1) \, \nu(z_2) \, \mu(z_3) \, h_1(z_4) \ra.
\label{second}
\ee

The fields $\sigma,\mu$ and $\nu$ are the fields discussed earlier, while the field $h_1(z)$ is the boundary scaling field corresponding to the lattice observable $\delta(h_z-1)-P_1^{\rm op}$, i.e. the deviation of the probability of having a height 1 at $z$ from the value it would have if the boundary was fully open (and no other field inserted). The scaling fields $h_i(z)$ for all four height variables on an open or closed boundary are known exactly \cite{jeng2,piru2}. In particular the scaling field for the height 1 on an open boundary is proportional to the boundary stress-energy tensor\footnote{Boundary height variables can be given simple expressions in terms of symplectic fermions \cite{piru2}. However since then it has been proved in \cite{jpr} that the symplectic fermions cannot account for all aspects of the sandpile model.} \cite{piru2}
\be
h_1(z) = \Big({3 \over \pi} - {80 \over 3\pi^2} + {512 \over 9\pi^3}\Big) \; T(z).
\label{bh1}
\ee

As $T(z)=(L_{-2})\bi(z)$, the 4-point amplitude (\ref{second}) is computed simply by applying the appropriate differential operator on the 3-point correlator $\la \sigma(z_1) \, \nu(z_2) \, \mu(z_3)\ra = C z_{13}^{1/2} \, z_{12}^{-1/2} \, z_{23}^{-1/4}$. The result is, with $A_1$ the numerical factor in (\ref{bh1}),
\be
\la \sigma(z_1) \, \nu(z_2) \, \mu(z_3) \, h_1(z_4) \ra = -{A_1 z^{}_{23} \over 8 z^{}_{14} z^2_{24} z^2_{34}} \Big[z_{12} (z_{24} + 3z_{34}) + z_{23} z_{24}\Big] \; \la \sigma(z_1) \, \nu(z_2) \, \mu(z_3)\ra.
\ee
The expansion for $z_2 \to z_3$ and the OPE (\ref{numu}) yields the relations
\bea
&& \la \sigma(z_1) \xi_2(z_3) h_1(z_4) \ra = 0\,, \qquad
\la \sigma(z_1) \phi_2(z_2) h_1(z_4) \ra = 0, \\
\noalign{\medskip}
&& {1 \over 2\beta_2} C_{\nu\mu}^{\R_{2,1}} \, \la \sigma(z_1) \psi_2(z_3) h_1(z_4) \ra = 
A_1 C {z_{13} \over 2z^{}_{14} z_{34}^3}.
\eea
Because $\sigma$ has zero weight and since $h_1$ is a descendant of a zero weight field, the field $\psi_2$ behaves like a primary field in the 3-point amplitude. 

\begin{figure}[t]
\epsfig{file=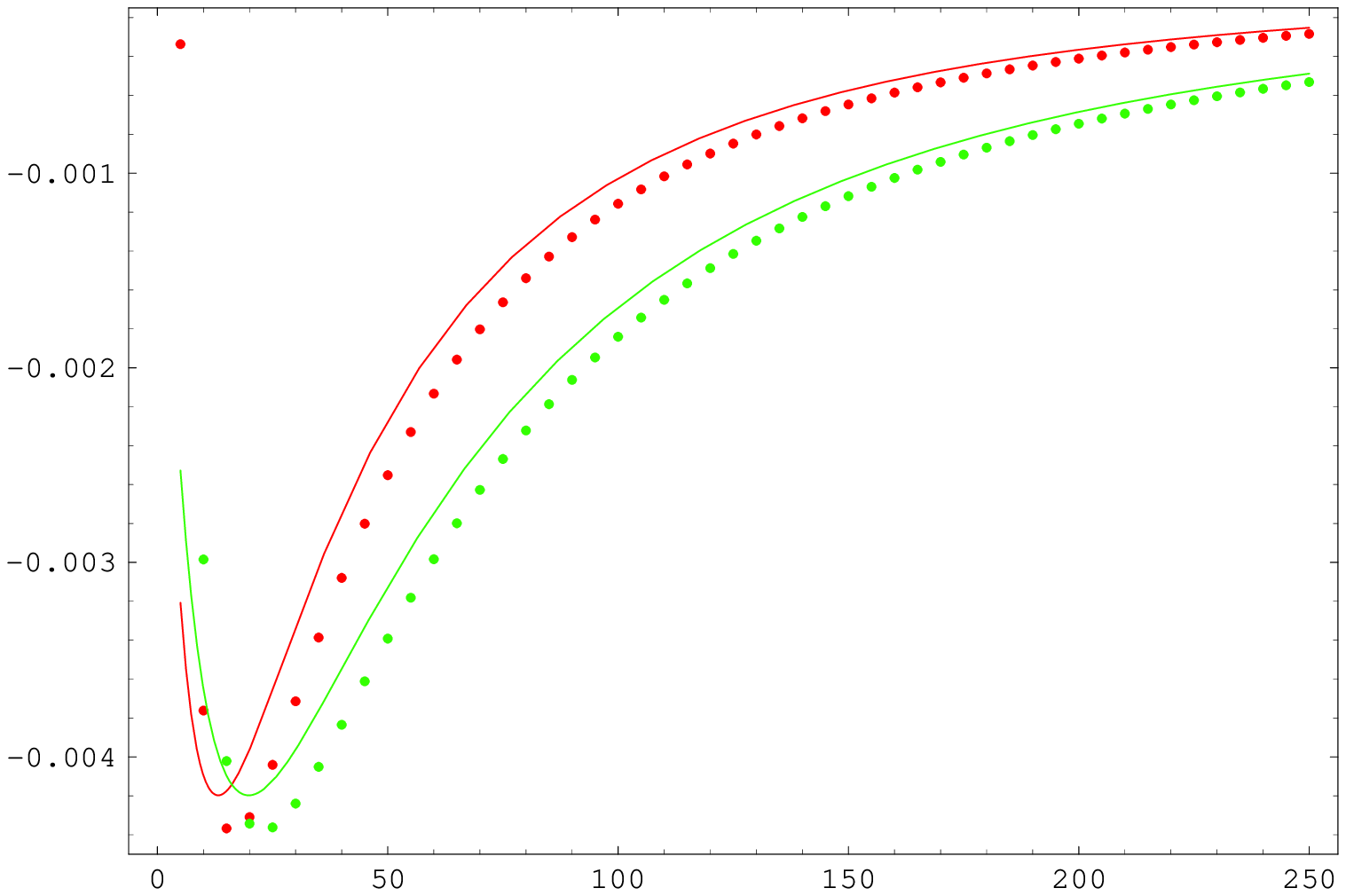,width = 7.5cm}
\hspace{.5cm}
\epsfig{file=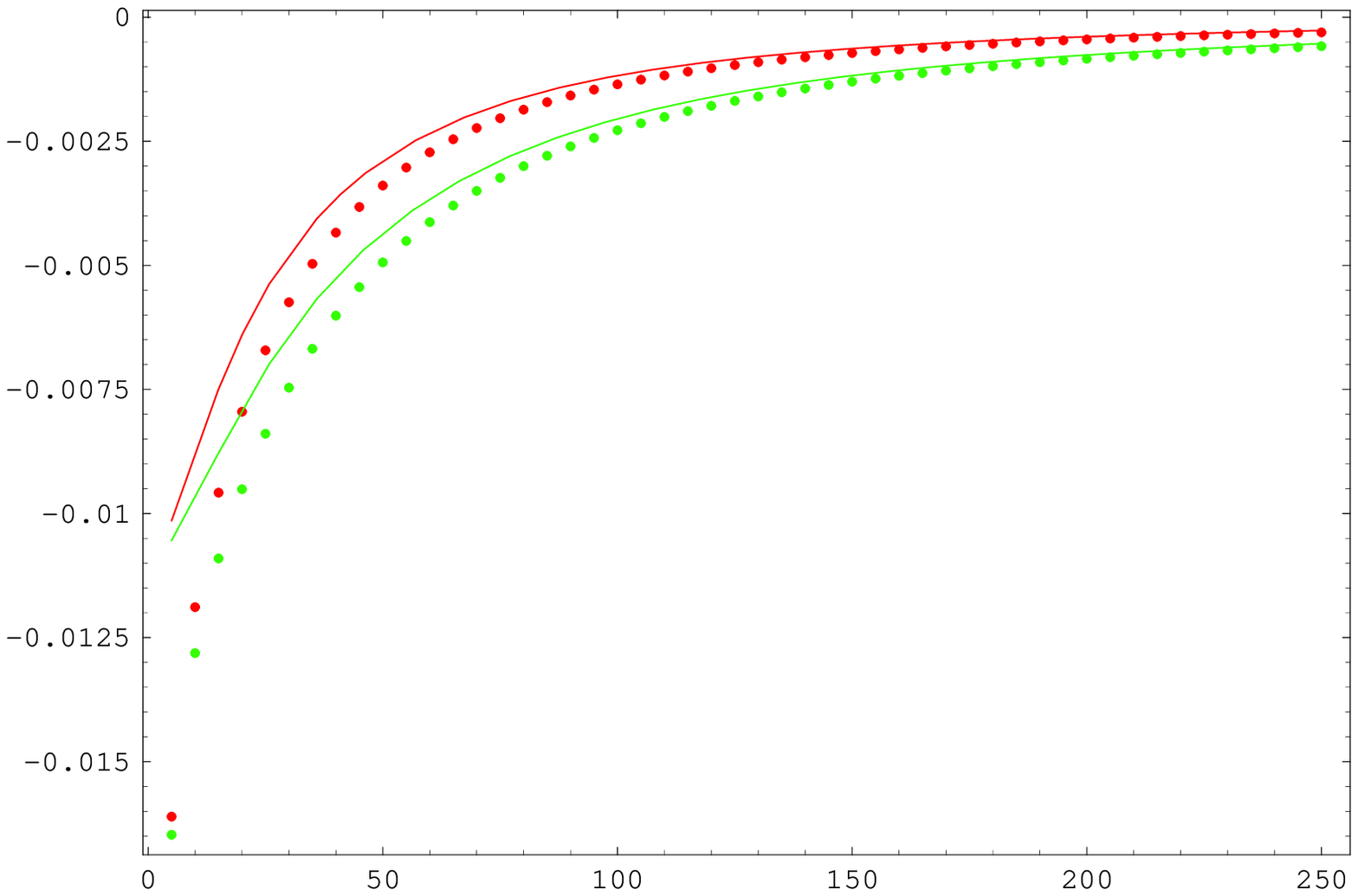,width = 7.5cm}
\caption{Comparison of the numerical (dots) and the conformal (solid) results for the probability difference (\ref{p1b}) for lengths $z_{21} = z_{32} = L$, taken to be equal to 20 (red) and 30 (green), and a varying position $z_4$. The left figure shows the results when the height 1 is at a distance $n$ to the left of the closed segment, so that $z_{43}=-n-2L$ with $n$ between 1 and 250. The right figure shows similar results when the height 1 is at a distance $n$ to the right of the segment of arrows, so that $z_{43}=n$, again running from 1 to 250. For convenience, both the numerical data and the conformal curves have been multiplied by a factor $n^2$, so that the discrepancy between the dots and the solid lines is actually much smaller than what it appears to be on the vertical scale.}
\end{figure}

Note that the ratio
\be
P_1(z_4|{\rm b.c.}) - P_1^{\rm op} = {\la \sigma(z_1) \, \nu(z_2) \, \mu(z_3) \, h_1(z_4) \ra \over \la \sigma(z_1) \, \nu(z_2) \, \mu(z_3)\ra} = -{A_1 z^{}_{23} \over 8 z^{}_{14} z^2_{24} z^2_{34}} \Big[z_{12} (z_{24} + 3z_{34}) + z_{23} z_{24}\Big]
\label{p1b}
\ee
is the probability that the height at the open boundary site $z_4$ is equal to 1, given the boundary conditions pictured in (\ref{second}), minus the same probability but with a fully open boundary ($P_1^{\rm op} \sim 0.104$ \cite{ivash}). This probability has been computed numerically for different positions of the various fields. The results are shown in Figure 2, and compare well with the conformal formula.

\subsection{\protect\boldmath Amplitude with $\R_{3,1}$} 

As a second example, in fact the only 4-point amplitude which involves the fusion to $\R_{3,1}$, we consider the sandpile model on the UHP with an open boundary, containing a first segment of right arrows, then a second interval of closed boundary sites, and finally a third segment of left arrows, as depicted here,
\be
\setlength{\unitlength}{1mm}
\begin{picture}(80,7)(0,-1)
\multiput(0,0)(2,0){8}{\line(1,0){1}}
\put(15,-1){\line(0,1){2}}\Blue
\put(15,0){\vector(1,0){3}}
\put(18,0){\vector(1,0){3}}
\put(21,0){\vector(1,0){3}}
\put(24,0){\vector(1,0){3}}
\put(27,0){\vector(1,0){3}}
\put(30,0){\line(1,0){2}}\Red
\put(32,0){\line(1,0){15}}\Black
\put(32,-1){\line(0,1){2}}
\put(47,-1){\line(0,1){2}}
\put(39.5,-2){\makebox(0,0)[c]{\small cl}}\Blue
\put(15,0){
\put(34,0){\line(-1,0){2}}
\put(37,0){\vector(-1,0){3}}
\put(40,0){\vector(-1,0){3}}
\put(43,0){\vector(-1,0){3}}
\put(46,0){\vector(-1,0){3}}
\put(49,0){\vector(-1,0){3}}\Black
\put(49,-1){\line(0,1){2}}
\multiput(49,0)(2,0){8}{\line(1,0){1}}
\put(56.5,-2.5){\makebox(0,0)[c]{\small op}}
}
\put(7.5,-2.5){\makebox(0,0)[c]{\small op}}
\put(15,3){\makebox(0,0)[c]{\small $z_1$}}
\put(32,3){\makebox(0,0)[c]{\small $z_2$}}
\put(47,3){\makebox(0,0)[c]{\small $z_3$}}
\put(64,3){\makebox(0,0)[c]{\small $z_4$}}
\end{picture}
\quad \sim \quad
\la \sigma(z_1) \, \nu(z_2) \, \nu(z_3) \, \s(z_4) \ra.
\label{third}
\ee

\smallskip
From the null descendant of $\sigma$ at level 3, $(L_{-1}^2 - 2L_{-2})L_{-1}\sigma=0$, one can find the general form of the 4-point function as the solution of a third order differential equation,
\be
\la \sigma(z_1) \, \nu(z_2) \, \nu(z_3) \, \sigma(z_4) \ra = z_{23}^{-3/4} \; \Big\{
\alpha\;\Big( {1 \over \sqrt{x}} - \sqrt{x}\Big) + \beta + \gamma\; \Big(\log{1-\sqrt{x} \over 1+\sqrt{x}} + 2\sqrt{x}\Big) \Big\}.
\ee
The OPE $\nu(z_2) \nu(z_3)$ should close on fields that interpolate between right arrows and left arrows (with a closed central site). If one postulates that, among these interpolating fields, the one with smallest dimension is the boundary condition changing field, then one can conclude that the OPE $\nu(z_2) \nu(z_3)$ cannot expand on fields with dimension lower than 1. This  readily implies $\beta=\gamma=0$, and we are left with
\be
\la \sigma(z_1) \, \nu(z_2) \, \nu(z_3) \, \sigma(z_4) \ra = \alpha \; z_{23}^{-3/4} \; {1-x \over \sqrt{x}}.
\label{snns}
\ee

To compare with the lattice results, we have set $z_{21}=z_{43}=L$, taken as before equal to 20, 30, 50 and 70, and $z_{32}=n$, varying between 1 and 150. Then $x$ is equal to $({L \over L+n})^2=t^2$ and the 4-point function reads
\be
\la \sigma(0) \, \nu(L) \, \nu(L+n) \, \sigma(2L+n) \ra = {\alpha \over n^{3/4}} \; {1-t^2 \over t}.
\label{fig2}
\ee
This function of $t$ (i.e. omitting the factor $n^{-3/4}$) has been plotted in Figure 3, along with the numerical values of the following lattice data
\be
n^{3/4} \; e^{2(4L+n)G/\pi} \; {\det \D^{\rm new} \over \det \Dop}.
\label{fig2latt}
\ee
Again the data collapse is clear and agrees very well with the result (\ref{fig2}) for a fitted value of $\alpha \simeq 1$. 

\begin{figure}[t]
\begin{center}
\epsfig{file=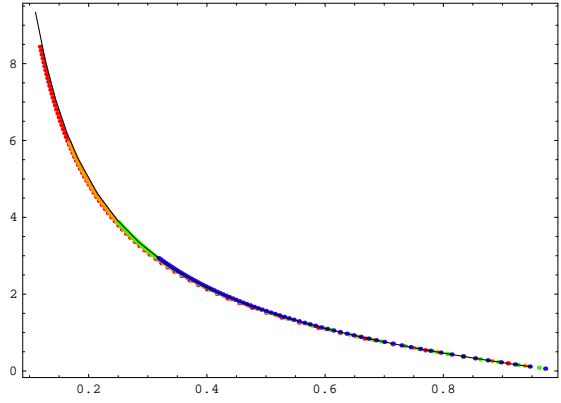,width = 7.5cm}
\end{center}
\caption{Comparison of numerical data against the conformal result for the situations pictured in (\ref{third}). The colour dots represent the lattice numerical results for (\ref{fig2latt}), for $n$ from 1 up to 150 and $L=20$ (red), 30 (orange), 50 (green) and 70 (blue). The horizontal coordinate is $t={L \over L+n}$.}
\end{figure}

As we did in the previous subsection, the structure of $\R_{3,1}$ can be confirmed. Matching the general OPE (where $\phi_3 = (L_{-1}^2 - 2L_{-2}^{}) \xi_3$ is the first singular descendant of $\xi_3$)
\bea
\nu(z) \nu(0) \egal C_{\nu\nu}^{\R_{3,1}} z^{1/4} \Big\{\xi_3(0) + {z \over 2} (L_{-1}\xi_3)(0) + {11 \over 24} z^2 (L_{-2}\xi_3)(0) - {3 \over 4\beta_3} z^2 \log{z} \, \phi_3(0)\nonumber\\
&& \hspace{3cm} -{3 \over 4\beta_3} z^2 \psi_3(0) + \ldots \Big\}
\eea
with the expansion of the 4-point function (\ref{snns}) yields
\bea
&& C_{\nu\nu}^{\R_{3,1}} \la \sigma(z_1) \xi_3(z_3) \sigma(z_4) \ra = \alpha \Big({1 \over z_{13}} + {1 \over z_{34}}\Big)\,, \\
&& C_{\nu\nu}^{\R_{3,1}} \la \sigma(z_1) (L_{-1}\xi_3)(z_3) \sigma(z_4) \ra = \alpha \Big({1 \over z_{13}^2} - {1 \over z_{34}^2}\Big)\,, \\
&& C_{\nu\nu}^{\R_{3,1}} \la \sigma(z_1) \phi_3(z_3) \sigma(z_4) \ra = 0\,, \\
&& C_{\nu\nu}^{\R_{3,1}} \la \sigma(z_1) \psi_3(z_3) \sigma(z_4) \ra = -{\alpha \beta_3 \over 18} \Big(-{2 \over z_{13}^3} + {3 \over z_{13}^2z_{34}^{}} + {3 \over z_{13}^{}z_{34}^2} - {2 \over z_{34}^3}\Big)\,.
\eea
These relations are compatible with the structure of $\R_{3,1}$. Indeed the second relation follows from the first one, and the fourth relation is compatible with the second one and $L_1\psi_3 = {\beta_3 \over 2} L_{-1}\xi_3$. The third equation follows from the first one and $\phi_3 = (L_{-1}^2 - 2L_{-2}^{})\xi_3$, \be
\la \sigma(z_1) \, \phi_3(z_3) \, \sigma(z_4) \ra = ({\cal L}_{-1}^2 - 2{\cal L}_{-2}^{}) \la \sigma(z_1) \, \xi_3(z_3) \, \sigma(z_4) \ra = \alpha ({\cal L}_{-1}^2 - 2{\cal L}_{-2}^{}) {z_{14} \over z_{13} z_{34}} = 0.
\ee
Again the decoupling of $\phi_3$ from this amplitude does not allow us to extract the value of the parameter $\beta_3$.  

\subsection{An example with dissipation} 

As a last example, we examine a situation on the UHP which involves boundary condition changes on a non-dissipative boundary. In conformal terms, and because the boundary remains non-dissipative, one has to add a dissipation field $\omega_N$ at infinity. This provides another opportunity to check the consistency of the conformal picture, and allows further checks on the field $\xi_3$ discussed just above. 

We consider the UHP with a closed boundary in which we insert a string of right arrows, immediately followed by a string of left arrows. Moreover the site at the junction between the two strings of arrows is taken to be closed. This situation is described in the scaling limit by,
\be
\setlength{\unitlength}{1mm}
\begin{picture}(80,5)(0,-1)\Red
\put(0,0){\line(1,0){15}}\Black
\put(15,-1){\line(0,1){2}}\Blue
\put(15,0){\vector(1,0){3}}
\put(18,0){\vector(1,0){3}}
\put(21,0){\vector(1,0){3}}
\put(24,0){\vector(1,0){3}}
\put(27,0){\vector(1,0){3}}
\put(30,0){\line(1,0){2}}
\put(34,0){\line(-1,0){2}}
\put(37,0){\vector(-1,0){3}}
\put(40,0){\vector(-1,0){3}}
\put(43,0){\vector(-1,0){3}}
\put(46,0){\vector(-1,0){3}}
\put(49,0){\vector(-1,0){3}}\Red
\put(32,0){\circle*{1.5}}\Black
\put(49,-1){\line(0,1){2}}\Red
\put(49,0){\line(1,0){26}}\Black
\put(75,0){\circle*{1.5}}
\put(75,0){\line(0,-1){2}}
\put(73,-2){\line(1,0){4}}
\put(73.5,-3){\line(1,0){3}}
\put(74,-4){\line(1,0){2}}
\put(32,-2.5){\makebox(0,0)[c]{\small cl}}
\put(75,3){\makebox(0,0)[c]{\small $\infty$}}
\put(7.5,-2.5){\makebox(0,0)[c]{\small cl}}
\put(62,-2.5){\makebox(0,0)[c]{\small cl}}
\put(15,3){\makebox(0,0)[c]{\small $z_1$}}
\put(32,3){\makebox(0,0)[c]{\small $z_2$}}
\put(49,3){\makebox(0,0)[c]{\small $z_3$}}
\end{picture}
\quad \sim \quad
\la \mu'(z_1) \, \xi_3(z_2) \, \mu'(z_3) \, \omega_N(\infty) \ra.
\label{fourth}
\ee

\noindent
We first consider this 4-point correlator with the dissipation field located at a finite position $z_4$, and then let $z_4 \to \infty$.

{}From the degeneracy of $\mu'$ at level two, the general 4-point function satisfies a second-order differential equation (even though $\omega_N$ is the logarithmic partner of the identity, this equation is homogeneous because the 3-point function $\la \mu'(z_1) \, \xi_3(z_2) \, \mu'(z_3) \ra$ vanishes in the absence of dissipation), whose general solution reads
\be
\la \mu'(z_1) \, \xi_3(z_2) \, \mu'(z_3) \, \omega_N(z_4) \ra = {z_{31}^{5/4} \over z_{21} z_{32}} \, \sqrt{x(1-x)} \, \Big\{ \alpha + \beta \Big(\sqrt{x \over 1-x} + \arcsin{\sqrt{1-x}} \Big)\Big\}.
\ee
Since the configuration of arrows is symmetrical around the point $z_2$, one expects the corresponding amplitude to be invariant under the exchange $z_1 \leftrightarrow z_3$. As this transforms $x$ into $1-x$, the above amplitude cannot be invariant unless $\beta=0$, so that
\be
\la \mu'(z_1) \, \xi_3(z_2) \, \mu'(z_3) \, \omega_N(z_4) \ra = \alpha {z_{31}^{5/4} \over z_{21} z_{32}} \, \sqrt{x(1-x)}.
\label{conf}
\ee

In the limit $z_4 \to \infty$, in which $x={z_{12}z_{34} \over z_{13}z_{24}} \to t = {z_{12} \over z_{13}}$, it becomes
\be
\la \mu'(z_1) \, \xi_3(z_2) \, \mu'(z_3) \, \omega_N(\infty) \ra = {\alpha \over z_{21}^{3/4}} \, {t^{1/4} \over \sqrt{1-t}}.
\label{fig4}
\ee

\begin{figure}[t]
\begin{center}
\epsfig{file=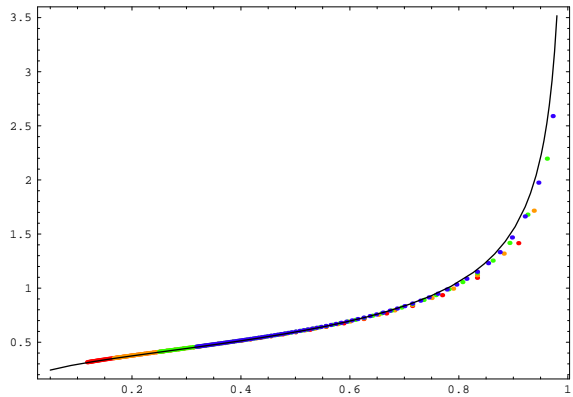,width = 7.5cm}
\end{center}
\caption{Comparison of numerical results against the conformal result for the situation depicted in (\ref{fourth}). The colour dots represent the lattice numerical results for (\ref{fig3latt}), and for $n$ from 1 up to 150 and $L=20$ (red), 30 (orange), 50 (green) and 70 (blue). The solid curve represents the function of $t$ in (\ref{fig4}). The abscissa is $t={L \over L+n}$.}
\end{figure}

On the lattice, we have computed numerically the effect caused by the insertion of arrows (the ratio of determinants is finite). We have taken the lengths of the two strings of arrows to be $L=z_{21}$ and $n=z_{32}$ respectively. As before the value of $L$ is fixed to $L=20,\,30,\,50$ and 70, and for each value of $L$, $n$ ranges between 1 and  150. The corresponding ratio of determinants have been divided by the usual exponential factor, here equal to $e^{-2(L+n)G/\pi}$. In addition, the previous conformal result suggests to first multiply the data by $z_{21}^{3/4}=L^{3/4}$ and then to plot them against the variable $t={L \over L+n}$. Figure 4 shows the comparison between the function of $t$ in the conformal result above, and the numerical values of
\be
L^{3/4} \; e^{2(L+n)G/\pi} \; {\det \D^{\rm new} \over \det \Dcl}.
\label{fig3latt}
\ee
The plots show an excellent agreement, for a fitted value of the coefficient $\alpha \simeq {1 \over 2}$. 

As a last remark, one should mention that $\phi_3$, the singular descendant of $\xi_3$ at level 2, also decouples here. This readily follows from the fact that the function (\ref{conf}) is in the kernel of the second order operator implementing the level 2 degeneracy condition on the correlators,
\be
\la \mu'(z_1) \, \phi_3(z_2) \, \mu'(z_3) \, \omega_N(z_4) \ra = ({\cal L}_{-1}^2 - 2{\cal L}_{-2}^{}) \la \mu'(z_1) \, \xi_3(z_2) \, \mu'(z_3) \, \omega_N(z_4) \ra = 0.
\ee


\section{Conclusions and perspective}

To summarize, we have defined, in terms of the spanning tree variables, two new boundary conditions in the two-dimensional Abelian sandpile model, and we have explored their nature in terms of a logarithmic CFT with central charge $c=-2$. Together with the well-known open and closed boundary conditions, they lead to seven new boundary condition changing fields $\phi^{\alpha,\beta}$, with scaling dimensions in the set $\{-{1 \over 8},\, 0,\, {3 \over 8},\, 1\}$. We have examined many 3- and 4-point amplitudes within the conformal setting, and have found a full agreement with the corresponding lattice data. This brings further support to the consistency and relevance of the conformal description, and adds new entries in this description.

It is worth stressing again the peculiarities and unusual features that these new boundary conditions have.
\begin{enumerate}
\item The arrow boundary conditions carry an intrinsic orientation. To our knowledge, this is the first instance of oriented boundary conditions. A direct consequence of this is that the boundary condition changing fields which involve arrows have a vanishing two-point function, or, formulated in another way, the metric on the space of states is off-diagonal in the representation basis. 
\item They cannot be uniformly imposed on a whole boundary. The reason for this is very clear in the sandpile model. The recurrent configurations are in one-to-one correspondence with spanning trees. Since a uniform arrow boundary condition on a boundary introduces a loop, it cannot be part of a spanning tree and therefore does not correspond to an allowed height configuration in the sandpile. In the conformal description however, this is a rather strange and new situation. The full implications of this from the general point of view of Boundary CFT need to be clarified.
\item All boundary condition changing fields are primary, but two of them, $\xi^{}_2 = \phi^{{\rm op},\lar}$ and $\xi'_2 = \phi^{\to\stackrel{\rm op}{,}\leftarrow}$, belong to an indecomposable representation $\R_{2,1}$ with rank 2 Jordan cells. The physical meaning of this and the physical interpretation of the lowest logarithmic partners (the $\psi^{}_2$ and $\psi'_2$ fields) remain to be understood.
\item A third boundary condition changing field, namely $\xi_3 = \phi^{\to\stackrel{\rm cl}{,}\leftarrow}$, is the lowest lying state of an indecomposable representation $\R_{3,1}$. In $\R_{3,1}$ this field is degenerate at level 5 but possesses a singular descendant at level 2. Even though this singular descendant has been seen to decouple in a number of amplitudes, we have no conclusive argument that it is actually null, and so we leave this question open. The same question as in point 3 regarding the physical interpretation of the field $\psi_3$ remains.
\end{enumerate}

Besides the questions raised above, there is clearly a number of other problems that need be answered before we can pretend to understand the sandpile model with boundaries. Among the most pressing and important ones, one can first mention the classification of all the observables which either preserve or interpolate between the four boundary conditions discussed here. Then one should also classify all conformally invariant boundary conditions present in the sandpile model, find their conformal description and determine the corresponding spectra of boundary fields. This is obviously much more challenging since an infinite number of boundary conditions are expected.


\section*{Acknowledgments}
It is a great pleasure to thank Vyatcheslav Priezzhev for precious discussions and suggestions during the early stages of this work. I am also grateful to Matthias Gaberdiel and Jorgen Rasmussen for useful discussions, patient explanations and for valuable comments on the manuscript. This work is partially supported by  the Belgian Internuniversity Attraction Poles Program P6/02. The author is a Research Associate of the Belgian National Fund for Scientific Research (FNRS).



\begin{thebibliography}{99}

\bibitem{rosa} L. Rozansky and H. Saleur, Nucl. Phys. {\bf B 376}, 461 (1992). 

\bibitem{sal} H. Saleur, Nucl. Phys. {\bf B 382}, 486 (1992).

\bibitem{gur} V. Gurarie, Nucl. Phys. {\bf B 410}, 535 (1993).

\bibitem{flo} M.A. Flohr, Int. J. Mod. Phys. {\bf A 18}, 4497 (2003).

\bibitem{gab} M.R. Gaberdiel, Int. J. Mod. Phys. {\bf A 18}, 4593 (2003).

\bibitem{maru} S. Mahieu and P. Ruelle, Phys. Rev. {\bf E 64}, 066130 (2001).

\bibitem{gulu} V. Gurarie and A.W.W. Ludwig, J. Phys. {\bf A 35}, L377 (2002); {\it Conformal Field Theory at central charge c=0 and Two-Dimensional Critical Systems with Quenched Disorder}, {\tt hep-th/0409105}. 

\bibitem{ru} P. Ruelle, Phys. Lett. {\bf B 539}, 172 (2002).

\bibitem{dgnpr} J. de Gier, B. Nienhuis, P.A. Pearce and V. Rittenberg, J. Stat. Phys. {\bf 114}, 1 (2004).

\bibitem{jeng1} M. Jeng, Phys. Rev. {\bf E 69}, 051302 (2004).

\bibitem{piru1} G. Piroux and P. Ruelle, J. Stat Mech. P10005 (2004).

\bibitem{jeng2} M. Jeng, Phys. Rev. {\bf E 71}, 036153 (2005).

\bibitem{jeng3} M. Jeng, Phys. Rev. {\bf E 71}, 016140 (2005).

\bibitem{piru2} G. Piroux and P. Ruelle, J. Phys. A: Math. Gen. {\bf 38}, 1451 (2005).

\bibitem{piru3} G. Piroux and P. Ruelle, Phys. Lett. B{\bf 607}, 188 (2005).

\bibitem{marr} S. Moghimi-Araghi, M.A. Rajabpour and S. Rouhani, Nucl. Phys. B 718,
362 (2005).

\bibitem{flomu} M.A. Flohr and A. M\"uller-Lohmann, J. Stat. Mech. P12006 (2005); J. Stat. Mech. P04002 (2006). 

\bibitem{iprh} N.Sh. Izmailian, V.B. Priezzhev, P. Ruelle and C.-K. Hu, Phys. Rev.
Lett. {\bf 95}, 260602 (2005).

\bibitem{jpr} M. Jeng, G. Piroux and P. Ruelle, J. Stat. Mech. P10015 (2006).

\bibitem{prz} P.A. Pearce, J. Rasmussen and J.-B. Zuber, J. Stat. Mech. P11017 (2006).

\bibitem{pr} P.A. Pearce and J. Rasmussen, J. Stat. Mech. P02015 (2007).

\bibitem{resa} N. Read and H. Saleur, Nucl. Phys. {\bf B 777}, 316 (2007).

\bibitem{ivash} E.V. Ivashkevich, J. Phys. {\bf A 32}, 1691 (1999).

\bibitem{sadu} H. Saleur and B. Duplantier, Phys. Rev. Lett. 58, 22 (1987) 2325.

\bibitem{lpsa} R. Langlands, P. Pouliot and Y. Saint-Aubin, Bull. Am. Math. Soc. {\bf 30}, 1 (1994).

\bibitem{car} J.L. Cardy, J. Phys. A: Math. Gen. {\bf 25}, L201 (1992).

\bibitem{wat} G.M.T. Watts, J. Phys. A: Math. Gen. {\bf 29}, L363 (1996). 

\bibitem{pr2} J. Rasmussen and P.A. Pearce, {\it Fusion Algebra of Critical Percolation}, {\tt arXiv:0706.2716}.

\bibitem{pr3} J. Rasmussen and P.A. Pearce, {\it Fusion Algebras of Logarithmic Critical Models}, {\tt arXiv:0707.3189}.

\bibitem{gk} M. R. Gaberdiel and H. G. Kausch, Phys. Lett. {\bf B 386}, 131 (1996); Nucl.
Phys. {\bf B 538}, 631 (1999).

\bibitem{garu} M.R. Gaberdiel and I. Runkel, J. Phys. A: Math. Gen. {\bf 39}, 14745 (2006).

\bibitem{gk2} M. R. Gaberdiel and H. G. Kausch, Nucl. Phys. {\bf B 477}, 293 (1996).

\bibitem{roh} F. Rohsiepe, {\it On reducible but indecomposable representations of the Virasoro algebra}, {\tt hep-th/9611160}.

\bibitem{eberflo} H. Eberle and M. Flohr, J. Phys. A: Math. Gen. {\bf 39}, 15245 (2006).

\bibitem{ipr} N.Sh. Izmailian, V.B. Priezzhev and P. Ruelle, SIGMA {\bf 3} (2007), 001.

\bibitem{btw} P. Bak, C. Tang and K. Wiesenfeld, Phys. Rev. Lett. {\bf 59}, 381 (1987).

\bibitem{dd} D. Dhar, Physica {\bf A 369}, 29 (2006).

\bibitem{priez} V.B. Priezzhev, Sov. Phys. Usp. {\bf 28} (12), 1125 (1985).

\bibitem{fishar} M.E. Fisher and R.E. Hartwig, Adv. Chem. Phys. {\bf 15}, 333 (1968).

\bibitem{ehrsi} T. Ehrhardt and B. Silbermann, J. Funct. Anal. {\bf 148}, 229 (1997).

\bibitem{widom} H. Widom, Amer. J. Math. {\bf 95}, 333 (1973).

\bibitem{feifu} B.L. Feigin and D.B. Fuchs, Funct. Anal. Appl. {\bf 17}, 241 (1983).







\end{thebibliography}
\end{document}